\documentclass[12pt]{iopart}
\usepackage{graphicx}% Include figure files

\begin{document}

%Transport and reactions of energetic particles in low-pressure, moderate-current discharges in H$_{2}

\review[Kinetics of discharges in H$_2$]{Collisional kinetics of non-uniform electric field, low-pressure, direct-current discharges in H$_{2}$}

\author{A. V. Phelps}
\address{JILA, National Institute of Standards and Technology and University of Colorado, Boulder, Colorado 80309-0440, USA} \ead{avp@jila.colorado.edu}
\date{\today}

\begin{abstract}

A model of the collisional kinetics of energetic hydrogen atoms, molecules, and ions in pure H$_2$ discharges is used to predict H$_\alpha$ emission profiles and spatial distributions of emission from the cathode regions of low-pressure, weakly-ionized discharges for comparison with a wide variety of experiments. Positive and negative ion energy distributions are also predicted. The model developed for spatially uniform electric fields and current densities less than $10^{-3}$ A/m$^2$ is extended to non-uniform electric fields, current densities of $10^{3}$ A/m$^2$, and electric field to gas density ratios $E/N = 1.3$ MTd at 0.002 to 5 Torr pressure. (1 Td = $10^{-21}$ V m$^2$ and 1 Torr = 133 Pa) The observed far-wing Doppler broadening and spatial distribution of the H$_\alpha$ emission is consistent with reactions among H$^+$, H$_2^+$, H$_3^+$, and H$^-$ ions, fast H atoms, and fast H$_2$ molecules, and with reflection, excitation, and attachment to fast H atoms at surfaces. The H$_\alpha$ excitation and H$^-$ formation occur principally by collisions of fast H, fast H$_2$, and H$^+$ with H$_2$. Model simplifications include using a one-dimensional geometry, a multi-beam transport model, and the average cathode-fall electric field. The H$_\alpha$ emission is linear with current density over eight orders of magnitude. The calculated ion energy distributions agree satisfactorily with experiment for H$_2^+$ and H$_3^+$, but are only in qualitative agreement for H$^+$ and H$^-$. The experiments successfully modelled range from short-gap, parallel-plane glow discharges to beam-like, electrostatic-confinement discharges.

\end{abstract}
\submitto{PSST}
\maketitle

\section{Introduction}
\label{sect:intro}

The objective of this paper is to test the ability of a previously developed model \cite{MODEL} of the kinetics of energetic ions, atoms, and molecules in low current discharges to explain quantitatively experimental observations of H$_\alpha$ emission and ion energy distributions from the cathode regions of a wide variety of low-pressure, moderate-current discharges in pure hydrogen.  In particular, calculated H$_\alpha$ Doppler broadened profiles, spatial distributions of spectrally integrated emission, and relative mass-identified ion flux energy distributions are reviewed and compared with experiment. The ability of the model to predict the observed emission in the far wings of the H$_\alpha$ profile is tested over wide ranges of current density and electric field to gas density $E/N$. The model utilizes cross sections and reaction pathways for various species in the production of fast H(n=3) atoms established for uniform electric fields in \cite{MODEL}, \cite{SPATIAL}, and \cite{DOPPLER}.  The model is extended to approximately describe the highly non-equilibrium behavior of the electrons in the spatially varying electric fields of the cathode fall and the negative glow. The production, transport, and H$_\alpha$ excitation by negative ions are estimated. The present steady-state model does not include radio frequency discharges, although many features of the observed H$_\alpha$ emission and the model are common to dc and rf discharges.

The observation of ``excess broadening'' in the wings of the H$_\alpha$ lines emitted by dc glow discharges in H$_2$ began many years ago. Sternberg and coworkers \cite{PRE77} found asymmetrical far-wing Doppler broadening of the Balmer lines from magnetron type discharges. Li-Ayers and Benesch \cite{BEN84} found that the magnitude of the far-wing profiles varied with cathode material as expected for reflection of the energetic particles responsible for excitation. May \cite{MAY85} used optogalvanic techniques to measure a Doppler broadened profile with a voltage dependent width for H(n=2) atoms in the cathode region of a Ne-H$_2$ glow discharge. These early observations illustrate the wide variety of laboratory experiments in which excess broadening is observed. More examples are cited in papers of this series \cite{PET92,MODEL,SPATIAL,DOPPLER}. The limited number of comparisons in the present paper are chosen to illustrate various aspects of the model for pure H$_2$, i.e., a review of every publication on this topic is not attempted.  This is an exploratory paper with a very much simplified discharge model that allows application to a wide range of experiments. The model will eventually need a better numerical treatment with more realistic geometries and using realistic differential collision cross sections. Thus far, only results using very approximate differential cross sections are available \cite{PET05,CVE05}.

The paper presents the simplest comparisons of calculations and experiment first, with the more approximate and less obvious comparisons later. Because of the complexity of the kinetics model, it is strongly recommended that readers first study the published model and experiments for uniform electric fields \cite{MODEL}. The paper begins with a discussion of representative electrode configurations and assumed electric-field distributions in \sref{sec:survey}. The range of experimental parameters covered is then reviewed. The required extensions of the kinetics model of \cite{MODEL} are summarized in \sref{sec:kinetics}.  Calculated and measured spatial and spectral distributions of emission and ion energy distributions are compared for planar cathodes in \sref{sec:planar} and for hollow or grid cathodes in \sref{sec:hollow}.  \Sref{sec:other} briefly discusses some related discharges that are not analyzed. The Appendices contain newly recommended cross sections, a discussion of relevant space charge effects, and an improved empirical description of hydrogen atom backscattering from surfaces.

\section{Discharge geometries and parameters} \label{sec:survey}

%figure 1

Because of the wide variety of electrode geometries considered, \fref{fig:electrodes} shows a schematic of the adopted one-dimensional planar-electrode model superimposed on representations of actual electrodes.
For the purposes of this paper the negative glow and the Faraday dark space are lumped together as a region of zero electric field \cite{DRU40}.  This figure shows the approximate electrode configuration including a planar cathode and anode \cite{BAR90,GAN91}, a planar cathode and a ring anode \cite{BAR90}, a planar cathode and a hollow anode \cite{CVE05}, a hollow cathode(s) and ring anode(s) \cite{LAV95,KIP08}, or a cathode and anode constructed of wire grids \cite{BOR09}.  The model places a planar anode at some representative position between the entrance to the hollow anode or the plane of the ring anode and the wall beyond these positions.  In the hollow anode case \cite{CVE05}, the emission data show considerable penetration of the discharge into the hollow anode. This also occurs for the wire electrodes \cite{BOR09}. Obviously, these geometrical approximations can influence the calculated emission, and eventually the correct geometry should be included in the models.  Electron backscattering from the anode \cite{PHE87} is neglected.

%figure 2

Having related the planar electrodes of the model to the physical electrodes, the approximate electric fields used in the model are discussed. The model replaces the spatially varying of the electric field with a constant one-dimensional electric field in the cathode fall region, and zero electric field in the negative glow and Faraday dark space \cite{DRU40}.  The constant electric field segments are estimates of the spatial average of the actual fields. \Fref{fig:electricfield} shows examples of the measured \cite{CVE05,BAR90,GAN91} and assumed electric-field $E$ to gas density $N$ ratios. (1 Td = $10^{-21}$ V m$^2$). In cases where the field was not measured, e.g., \cite{LAV95,KIP08}, the calculation uses the authors' discharge voltage divided by a measured or scaled gas pressure times cathode fall thickness \cite{FRA56}.

%figure3

A very wide range of discharge parameters has been \cite{MODEL,SPATIAL,DOPPLER} and will be covered. \Fref{fig:eonvscurrent} shows this range of conditions as a plot of average $E/N$ versus average discharge current densities. The range of pressures and distances is relatively limited ($0.2 < pd < 3$ Torr cm) and a three (or more) dimensional plot is not shown. The upward pointing (blue) triangles represent the low-current, low-pressure experiments from which the spatial scans of H$_\alpha$ emission of \cite{SPATIAL} were selected and used primarily to test the reaction kinetics and absolute values predicted by the model. The group of downward-pointing (green) triangles represent the data sets from which the H$_\alpha$ Doppler profiles of \cite{DOPPLER} were selected and used to test the velocity distributions of the emitting species and the absolute intensity predictions of the model.  These experiments were conducted at current densities of $< 10^{-2}$ A/m$^2$ for which the electric field is spatially uniform.  For these low current conditions, the spectrally integrated magnitudes of the H$_\alpha$ signal were found experimentally to be a linear function of the discharge current as indicated by dotted line at the low current densities \cite{SPATIAL,JEL85}. The tests of linearity for higher current densities \cite{JEL92,ADA03} are shown by the dotted lines and are discussed below.

The dashed line of \fref{fig:eonvscurrent} shows the approximate upper limit to the $E/N$ for which space-charge effects can be neglected and the electric field is spatially uniform. See \ref{sec:spacecharge} for the derivation of this condition.  For sufficiently large electrode separations and experimental current densities, i.e., to the right of the solid line of \fref{fig:eonvscurrent}, one expects a fully developed cathode fall \cite{DRU40}. The present paper includes the modelling of several such experiments \cite{CVE05,BAR90,KIP08,LAV93}. The parameters for experiments with measured spatial dependencies of the electric field \cite{CVE05,GAN91,BAR91} are shown in \fref{fig:electricfield}.  For the hollow-cathode experiments \cite{KIP08,LAV93} there is considerable uncertainty in the estimates of the spatial distributions of the both the electric field and the current. The averages of the $E/N$ between the cathode and anode for the experiments of Boris et al \cite{BOR09} and \sref{subsec:boris} are too high to plot in \fref{fig:electricfield}.

These analyses neglect the build up of atomic hydrogen as the result of dissociation of H$_2$ by the discharge and the resultant increase in, for example, symmetric and asymmetric charge transfer collisions of H$^+$ and H$_2^+$ with H atoms. Measurements of fractional H atom concentrations in rf discharges at input power levels comparable with those discussed in \sref{sec:planar} show hydrogen ground-state atom densities as high as several percent \cite{TER94},  These densities are dependent on highly variable surface recombination probabilities for H atoms, especially for Cu \cite{GAN94}. It is estimated that $\geq 1\%$ of H(1s) atoms are required for the symmetric charge transfer collision of H$^+$ with H(1s) to noticeably influence the present calculated results.  Changes in the H$_2$ translational temperature caused by the discharge are neglected because of the high kinetic energies of the relevant collisions.  Although the effects of increases in the H$_2$ vibrational temperature on electron attachment to H$_2$ can be large \cite{ALL78}, their effects are estimated to be small in this paper. These calculations adopt the custom of all the authors cited of neglecting possible changes in H$_2$ density resulting from temperature increases caused by the discharge.

\section{Kinetics model} \label{sec:kinetics}

This section is concerned with kinetic processes that were not included in the earlier versions of the model \cite{MODEL,SPATIAL,DOPPLER}. Because the experiments considered in this paper include observations from adjacent regions of high and low electric field at low gas densities, it is necessary to apply a non-local electric field model to the electrons.  This has been done in an approximate way by extending the multi-beam model for ions of \cite{MODEL} to the electrons.  Direct comparison of the electron model with experimental results are discussed in \cite{TRANS}.  The principle simplification from the form of equation (6) of \cite{MODEL} is that the energy loss per elastic scattering collision, given by equation (7), is small enough so that it is included in an energy loss function L$_e^{in}(\epsilon)$ calculated from the sum of the products of inelastic-energy thresholds and angular-integrated, excitation and ionization cross sections.  Ionization is assumed to produce a new electron at zero energy and an electron that has lost the ionization energy, but is not scattered in angle.  Electron-excited H(n=3) atoms are assumed to have low enough velocities so that they radiate without changing position.

%figure 4

The earlier model \cite{MODEL} has been extended to include the formation and loss of H$^-$, as illustrated in \fref{fig:kinetics}.  The cross sections for collisions of H$^-$ with H$_2$ are reviewed in \ref{sec:hminusxsect}. The electron attachment cross section is a fit to the high energy portion of the recommendation of Joon \etal \cite{JOO08} and is usually negligible for the high mean energy electrons considered even when allowances are made for possible dissociative attachment to vibrationally excited H$_2$ molecules. The estimated H$^-$ production at surfaces is typical of the experimental results of \cite{VER80,MAA96,WUR97}.  It is expressed as 0.04 H$^-$ per reflected fast H atom leaving the cathode and is assumed to be independent of energy.  In addition to collisional detachment \cite{PHE90}, the H$^-$ ions are assumed to be destroyed on collision with any surface.  However, recent experiments with a graphite cathode \cite{SCH09} raise the possibility of a significant energy dependence of the H$^-$ yield and of the role of adsorbed H$_2$.

Next, the process of H(n=3) production at surfaces is reconsidered. As shown in \cite{MODEL}, this process is not expected to be observed at moderate H$_2$ pressures.  However, it will be found to be important at very low pressures and observations near a surface \cite{KIP09}.  On the basis of experiments with hydrogen and deuterium ions for various surfaces \cite{LEU78,TAN95}, the probability of H(n=3) excitation per reflected H atom is assumed to be 0.1 at high energies, with the measured low velocity cutoff \cite{TAN95}.

A serious problem in the extension of the model of \cite{MODEL} to glow discharges is the treatment of radial losses, particularly in the negative glow-Faraday dark space \cite{DRU40} in axially extended geometries.  In these regions, fast H(n=3) atoms are produced by fast H atoms previously reflected from the cathode, while slow H(n=3) atoms are excited by electrons accelerated through the cathode fall.  The problem of charged-particle loss from these regions has been treated recently by Donk\'o \etal \cite{DON06}.  Based on the unpublished measurements of \cite{TRANS}, the calculations of the present paper are made assuming that electrons are scattered through a large angle and are lost to the wall at a rate determined by the momentum-transfer cross section.  The assumption of a concurrent ion loss resulting from the action of the ambipolar electric fields, created by the escaping electrons, primarily affects the low energy ions and has little effect on H(n=3) excitation.  An empirical alternative to loss by scattering is the loss of energy by the electrons, e.g., in \sref{subsec:cvet} a factor of 4 increase in the continuous energy-loss function of \cite{PHEWEB} would be required to fit the observed electron attenuation when large-angle scattering is neglected. Thus, it will be found that the apparent attenuation of the electron-excited portion (``core'') of the H$_\alpha$ profile depends primarily on the electron loss. The attenuation of the wings depends on the initial diffuse emission of the reflected fast H atoms \cite{MODEL}, on the subsequent radial scattering of the surviving fast H atom, and on the collisional energy loss discussed in \cite{MODEL}. For a fixed pressure and discharge geometry, the relative magnitude of the cathode-leaving wing component of the H$_\alpha$ profile is varied by adjusting the fast atom reflection coefficient at the cathode. Because of the multi-step nature of the H(n=3) excitation chain \cite{MODEL}, the cathode-approaching wing of the H$_\alpha$ profile generated in the cathode fall increases rapidly relative to the core as the thickness of the cathode fall increases.

The flux of positive ions, mostly H$_3^+$, entering the cathode fall from the negative glow \cite{DRU40} is effectively limited by the radial flow to the wall of the electrons that produce the ions \cite{TRANS}. In the present model, this flux is treated as an adjustable parameter, although it can be estimated by using a self-consistent treatment of secondary electron emission at the cathode \cite{DRU40}. The model will show that the H$_3^+$ entering the cathode fall is partially converted to H$_2^+$ and H$^+$ by energetic collisions in the high electric field region.

\section{Experiments with planar cathodes}\label{sec:planar}

\subsection{Experiments of Ganguly and Garscadden}\label{subsec:ganguly}

%figure 5

Analysis of the experiments by Ganguly and Garscadden \cite{GAN91} considered in this section is the first step in testing the extension of the uniform electric field model \cite{MODEL} to higher current densities and higher $E/N$. Their discharge current is 5 mA or 1 mA/cm$^2$ for an electrode separation of $d= 6.5$ mm and diameter of 25.4 mm at a pressure of $p = 0.3$ Torr and a discharge voltage of $V = 6.5$ kV. The discharge is an obstructed discharge \cite{DRU40} in which the measured electric field $E$, as shown in \fref{fig:electricfield}, varied only about 30\% and the average $E/N$ is 55 kTd. Thus, the parameters for this experiment fall very close to the dashed curve of \fref{fig:eonvscurrent} marking the onset of space charge effects. 

\Fref{fig:ganguly}(a) compares the calculated H$_\alpha$ profile (curve) and measured relative spectral distribution (points) for the H$_\alpha$ line as observed perpendicular to the electric fields (transverse to the discharge axis). The points are adjusted in magnitude to roughly fit the calculations. The vertical dashed (red) lines show the authors' choice of the dividing wavelength shift between large (wing) and small (core) Doppler shifts. The model profile is calculated assuming the earlier fit to the published fast heavy-particle reflection, angular distribution parameter for reflected H atoms ($b=0.6$), and reaction probabilities of \cite{MODEL} with the addition of the non-equilibrium behavior of electrons discussed in \sref{sec:kinetics}.  Thus, the adjustable parameter in the comparison in \fref{fig:ganguly}(a) is the magnitude scale factor applied to the experimental data. The calculated shape of the wings of the profile is good, but their magnitude is too small relative to the core of the line.  This discrepancy is in the direction expected because the model assumes that the effects of electrons backscattered from the anode are neglected \cite{PHE87}. Also note that the experiment shows departures from cylindrical symmetry \cite{GAN91} that may account for the observed asymmetries in the profile.

The dash-dot (black) curve of \fref{fig:ganguly}(b) shows the calculated spatial distribution of the integral of the H$_\alpha$ profile for $|\Delta\lambda| < 0.1$ nm, primarily caused by electron excitation. The dashed (red) curve shows the calculated sum of the contributions of fast excited H(n=3) atoms with $|\Delta\lambda| > 0.1$ nm, while the solid (blue) curve shows the total line intensity. The relative intensity scale is adjusted by eye to fit the calculations. The large decrease with position in the calculated emission in the core of the H$_\alpha$ line results from the radial loss of electrons as determined by the momentum-transfer cross section for electrons discussed in \sref{sec:kinetics} and in \cite{TRANS}.  As for experiments at low current densities in \cite{MODEL}, the present comparison with experiment assumes that the yield of reflected fast H atoms is that given by backscattering theory and experiment \cite{MODEL,ECK84,ECK85}.  The agreement between the calculated and measured spatial intensity variations is moderately good, although there are systematic discrepancies that are not understood.

An interesting aspect of the obstructed discharge used by Ganguly and Garscadden \cite{GAN91} is the small number of collisions of ions, fast atoms, and fast molecules with the background H$_2$ because of the relatively small $p d$, where $p$ is the pressure and $d$ is the electrode separation. Because of the multi-step kinetics leading to H$_\alpha$ production, there is relatively little excitation by fast H atoms or H$_2$ molecules approaching the cathode, and the data provide an enhanced opportunity to study the behavior of H atoms backscattered from the cathode. Evidence for this behavior is that the predicted, but unmeasured, axial H$_\alpha$ spectral profile is highly asymmetric for cathode surfaces with a high efficiency of fast atom reflection.

\subsection{Experiments of Barbeau and Jolly}\label{subsec:barbeau}

The extensive set of H$_\alpha$ profiles transverse to the electric field from the experiments of Barbeau and Jolly \cite{BAR90} offer us an opportunity to (1) test the present model of the spatial and pressure dependence of the relative emission in the wings of the H$_\alpha$ line, (2) possibly improve the model of the core of the H$_\alpha$ profiles in the limits of excitation by H atoms and of excitation by electrons, and (3) briefly review the evidence against a significant contribution to the H$_\alpha$ emission by H(n=3) atoms produced at the cathode surface. As shown in \fref{fig:electricfield}, Barbeau and Jolly \cite{BAR91} found that the electric field in the cathode fall at 0.6 Torr, 100 A/m$^2$, and 800 V decreases linearly with distance. Using their value for the product of pressure and the surprisingly-large cathode-fall thickness \cite{FRA56} of $\approx 0.7$ Torr cm and neglecting variations with current density \cite{FRA56}, the estimated average $E/N$ in the cathode fall varies from 2.1 and 4.6 kTd for pressures from 0.27 to 1 Torr. The use of a ring anode when making H$_\alpha$ profile observations along the axis of the discharge make the effective cathode-anode distance and anode area indefinite. The model results are therefore normalized to charged-particle fluxes at the cathode.

%Figure 6

The experiments of Barbeau and Jolly \cite{BAR90} are carried out at significantly higher pressures and larger distances than those of Ganguly and Garscadden \cite{GAN91}.  This requires that the model deal with the contribution of the negative glow region as well as that of the cathode fall. As an example of the predictions of the model, \fref{fig:barbeaufluxes} shows the calculated fluxes of ions, atoms, and fast molecules versus position for the lowest pressure experiment of Barbeau and Jolly \cite{BAR90}, i.e., $p=0.27$ Torr. The calculated fluxes are normalized to the electron flux at the cathode. In view of the relatively large radius of the discharge, it is assumed that all ions produced in the negative glow enter the cathode fall. Because of the low ion energies in the negative glow, the H$_2^+$ (dashed-orange curve) produced by electrons are rapidly converted to H$_3^+$ (dotted-green curve).  When the H$_3^+$ enter the high electric field of the cathode fall, many are converted to H$_2^+$ (dashed orange curve) and H$^+$ (dash-dot red curve).  The H$^-$ flux produced at the cathode during the reflection of H atoms (long-dash purple curve) is small and decays rapidly by collisional detachment. If one assumes that none of the ions produced between the ring anode and the cathode fall are injected into the cathode fall the ion and neutral species fluxes in the cathode fall region are significantly smaller than with ion injection. An example of the case of no ion injection is the lower dotted (green) curve for H$_3^+$.  Not surprisingly, the flux plots in the cathode fall region for no ion injection are qualitatively similar to those calculated for the high $E/N$, uniform electric-field case in figure 7(a) of \cite{MODEL}. Near the cathode, where most of the H$_\alpha$ is produced by the more energetic particles, the difference in fluxes for the two injection models is only $\sim 10\%$ for H$_3^+$ and $\sim 40\%$ for fast H$_2$.

%Figure 7

\Fref{fig:barbeauprofiles}(a) compares experimental and calculated examples of transverse profiles obtained by Barbeau and Jolly \cite{BAR90} at two positions along the axis of the discharge at a pressure of 1 Torr and a cathode-anode separation of 30 mm. These experimental profiles are normalized to the model, preserving the experimental relative magnitudes.  Because the contribution of electrons is small at 1 mm from the cathode, the transverse profile shown by the upper curve of \fref{fig:barbeauprofiles}(a), is expected to be representative of emission by H(n=3) atoms produced by dissociative excitation \cite{GED81} of H$_2$ by H atoms with energies in the 200 eV range. Empirically, the broad core for the H$_\alpha$ line can be approximated by the following normalized dispersion formula:
\begin{equation}\label{eq:dispersion}
I_d(\Delta\lambda) = (\pi w_d)^{-1}(1 + \Delta\lambda^2/w_d^2)^{-1} , \end{equation}
where $w_d = 0.07$ nm is the half-width-at-half-maximum (HWHM) of this profile \cite{AVP10i}. When using this relation, it is necessary to increase the assumed fraction of H atom excitation leading to emission in the core of the line to about 50\%, compared to the 10\% previously assumed \cite{MODEL}. In the lower curve of \fref{fig:barbeauprofiles}(a) for data taken at 14 mm from the cathode, a better fit is obtained using a Gaussian with a HWHM of 0.05 nm corresponding to about 3 eV atoms.  This change in the core behavior correlates with the calculated increase of electron induced dissociative excitation relative to excitation by fast H atoms at large distances from the cathode \cite{CVE05}.  Obviously, further work is required to test these suggested fits.

In the calculation of the axial profiles in \fref{fig:barbeauprofiles}(b), it is assumed that the angular distribution parameter in equation (8) of \cite{MODEL} decreased with pressure from $b=5$ for approaching H(n=3) atoms and $b=1$ for leaving H(n=3) atoms at 0.27 Torr to $b=0.2$ for all H(n=3) atoms at 1 Torr. Thus, good fits to experiment require one to assume that the angular distributions become significantly more isotropic as the pressure increases. These experimental axial profiles are independently normalized to the model.  An unknown parameter of the model is the effective distance into the negative glow that supplies ions, primarily H$_3^+$, to the cathode fall. One logical choice would be a distance determined by the balance between diffusion to the cathode-fall boundary and loss by radial diffusion by electron-ion recombination \cite{DRU40,DON06}. Empirically, the calculations assume negligible ion flux entering the cathode fall.  This assumption improves the agreement with the shape of the $\Delta\lambda < 0$ portion of the profiles of \fref{fig:barbeauprofiles}(b). For 0.27 Torr and negative $\Delta\lambda$ (H(n=3) approaching the cathode), the calculated excitation by fast H accounts for the -0.5 nm peak and 76\% of the area under the wings of the profile. The low-velocity peak at -0.2 nm is from excitation by H$_2$, which accounts for 19\% of the wings. Ions account for 5\%.  For 1 Torr, fast H accounts for 88\% of the area. Excitation by H$_2$ accounts for 6\%, and ions account for 6\%. Changing the model to assume that all ions formed in the negative glow enter the cathode fall raises the calculated profile by a factor of $\sim 2$, especially in the far negative wing.

The effects of heavy-particle-induced ionization and ion pair formation in collisions of the fast atoms, ions, and molecules with H$_2$ are modelled using the cross sections summarized in \ref{sec:hminusxsect}. The contributions of these processes to the spectral intensities shown in \fref{fig:barbeauprofiles}(b) at $\Delta\lambda <-0.2$ nm is $\sim 10\%$ at both 0.27 Torr and 1 Torr. For $\Delta\lambda >+0.2$ nm, these processes contribute $\sim 10\%$ at 0.27 Torr and $\sim 30\%$ at 1 Torr.

%Figure 8

The measurements of the relative intensities of the core and wing components of the transverse H$_\alpha$ profiles by Barbeau and Jolly \cite{BAR90} shown in \fref{fig:barbeaufraction} provide a further test of the ability of the model to predict the apparent attenuation of the H(n=3) atoms as one moves away from the cathode. This figure shows the measured and calculated fractions of the profiles that are emitted with wavelength shift greater than 0.1 nm, i.e., the fraction of the profile emitted by H(n=3) atoms with greater than 10.9 eV energy along the line of sight. This fraction is a measure of the relative importance of H$_\alpha$ excitation in the wings resulting from heavy-particle collisions with H$_2$ versus H$_\alpha$ excitation in the core by electrons and heavy particles. No corrections for backscattered optical radiation have been made \cite{ADA03}. The comparisons of calculations and experiment in \fref{fig:barbeaufraction} show that the model correctly predicts the change in the ratio of the core and wing components of the Doppler profiles with pressure.

In summary, the observations of Barbeau and Jolly \cite{BAR90} are quantitatively explained in terms of excitation of the wings of their H$_\alpha$ profiles by fast H atoms and H$_2$ molecules produced by ions approaching the cathode, plus excitation by fast H atoms produced on reflection at the cathode of ions, atoms, and molecules.

\subsection{Experiments of Konjevi\'c \etal}\label{subsec:cvet}

This section compares the predictions of the extended model with some of the many measurements of H$_\alpha$ emission from H$_2$ glow discharges by Konjevi\'c and co-workers \cite{CVE05,ADA03,KUR92,VID96,SIS05,CVE09}. Using high-resolution measurements of the Balmer series profiles, Cvetanovi\'c \etal \cite{CVE05} determined the electric field in the cathode region as shown in \fref{fig:electricfield} for their 0.29 Torr, 900 V discharge. The simplified model of this paper utilizes the average field. The approximation made in \sref{sec:survey} in which the hollow cylinder anode is replaced by a planar anode seems reasonable in view of the small thickness of the cathode fall (1.6 mm) compared to the internal diameters of the anode (5 or 8 mm) and the very small electric field expected in the negative glow \cite{DRU40}. The effective anode is placed at the maximum distance from the cathode for which data is shown. The distortion of the electric field in the portion of the negative glow located inside the hollow anode is neglected.  All of the model results of this section are normalized to unit total charged-particle flux at the cathode.  Because of the reported severe damage of the cathode by particle bombardment \cite{AVP10c}, model calculations multiply the reflected H atom flux from \ref{sec:backscattering} by a freely adjustable fitting parameter that varies from 0.1 to 0.7. The angular distribution parameters of the model \cite{MODEL} are near unity, i.e., $b=0.6$ for backscattered H atoms and $b=1$ for approaching H atoms, corresponding to near-cosine distributions.

Konjevi\'c and co-workers \cite{CVE05,ADA03,KUR92,VID96,SIS05,CVE09} have characterized the wings of their H$_\alpha$ profiles by fitting Gaussian profiles to the data and discussing the apparent ``temperatures'' of the excited H atoms.  It is important to keep in mind that these temperatures are measures of the energy spreads and of, roughly, the mean energies of the H(n=3) atoms with velocities directed along the line of observation. Contrary to a very recent publication \cite{LOU10}, these temperatures are not to be interpreted as indicating that the velocity distribution of the H(n=3) atoms is an isotropic Maxwellian. Furthermore, the reasonable fits of the smoothly-varying, displaced Gaussian functions over a significant range of relative spectral intensities and velocities effectively rule out the proposal \cite{LOU10} that mono-energetic and isotropic velocity distributions are a good approximation for the excited H atoms.

%Figure 9

\Fref{fig:cvetprofiles}(a) shows a comparison of calculated and measured Doppler profiles obtained by observation along the axis of the discharge, i.e., parallel to the electric field at the cathode. The published wavelength shifts have been reversed so as to conform to the notation in \cite{MODEL}, i.e., positive shifts represent H(n=3) atoms leaving the cathode and negative shifts represent H(n=3) atoms approaching the cathode. No corrections for optical reflection have been made \cite{DOPPLER,ADA03}. The calculated H$_\alpha$ profile is obtained by assuming that the reflected fast H atom flux is 40\% of that predicted by backscattering theory and experiment \cite{MODEL,ECK84,ECK85}. The shape of the calculated profile agrees well with the experiment. The shoulders predicted near $\Delta\lambda = -0.5$ nm and $-0.2$ nm are caused by excitation of H(n=3) atoms by fast H atoms and fast H$_2$ molecules, respectively.

\Fref{fig:cvetprofiles}(b) shows a comparison of calculated and measured Doppler profiles for observations transverse to the discharge axis in the negative glow region at the positions indicated in \fref{fig:cvetspatial}.  The points show the experimental data of Cvetanovi\'c \etal \cite{CVE05} plotted with the same scale factor for the two profiles.  The principle contributions to the calculated emission by H(n=3) atoms are from excitation by fast H and by fast H$_2$.  The contributions of H$_\alpha$ excitation caused by H$^+$, H$_2^+$, and H$_3^+$ approaching the cathode are small. The ``core'' component is mostly from excitation by electrons, plus some dissociative excitation by fast H and H$_2$.  As in \cite{MODEL}, the electron excitation portion is assumed to be 50\% dissociative excitation with an effective HWHM of 0.05 nm (3.5 eV) chosen to fit these experiments. The remaining dissociative electron excitation is assigned an effective full-width-at-half-maximum (FWHM) of 0.01 nm to simulate the fine structure splitting and Stark broadening \cite{VID96}.  As the point of observation is moved away from the cathode, there is a relative decrease in the excitation in the wings resulting from the loss of diffusely backscattered fast H atoms.  Here it is assumed that the effective reflection coefficients for the highly sputtered Cu cathode are 30\% of predicted values \cite{ECK84}.  As in the case of the upper transverse profile of \fref{fig:barbeauprofiles}(a), the assumption of Gaussian profiles for the broadening of the core does not yield good fits to the experimental profiles near $\Delta\lambda =\pm 0.2$ nm. The empirical fits of modified dispersion profiles discussed in \sref{subsec:barbeau} have not been tested against the present experiments.

%Figure 10

\Fref{fig:cvetspatial} shows a comparison of calculated and measured spatial distributions of the components of H$_\alpha$ emission obtained by fitting three Gaussians to the profiles measured transverse to the discharge axis at various points along the discharge. The area under their widest Gaussian, called G3 by these authors \cite{CVE05}, comes much closer to representing the contributions to the H$_\alpha$ profile resulting from fast H(n=3) atoms, some of which move along the discharge axis and have low transverse velocities, than does selection of high radial energy H(n=3) atoms used by other authors \cite{BAR90,GAN91}. The sum of the areas under their two narrower Gaussians (G1+G2) then represent the H$_\alpha$ excitation resulting from dissociative excitation of the target H$_2$ by electrons and by incident fast heavy particles. This sum is the core of the line in the present paper. In the model it is assumed that the reflection as H atoms is 70\% of the predicted values for Cu \cite{ECK84}, probably corresponding to a relatively undamaged cathode. This reflection parameter primarily determines the relative magnitude of the reflected- or leaving-atom component responsible for the wings of the profile in the negative glow, i.e., the (red) curve of \fref{fig:cvetspatial}.  As pointed out by the model of Cvetanovi\'c \etal \cite{CVE05,CVE09}, these data show that the diffusely reflected fast H atoms from the cathode are lost much more rapidly than the beam-like electrons injected into the negative glow from the cathode fall.

The reasons for the discrepancy between the model and experiment for the spatial dependence of emission within 1 mm of the 6 mm diameter Cu cathode are unknown. Similarly, the differences among the data \cite{CVE05} for the various lines of the Balmer series in this region are unexplained.  A structure in the total emission similar to that observed near the cathode can be obtained by subtracting the H$_\alpha$ emission attributed by the model to excitation by approaching heavy species, i.e., excitation by fast H and H$_2$, and shown by the dash-dot (green) curve in \fref{fig:cvetspatial}.  However, the analysis of \sref{subsec:lavrov} presents reasonably direct evidence of a significant contribution by both fast H and fast H$_2$ to excitation of H(n=3) atoms approaching the cathode.

The comparison of the predictions of the present model with the Doppler profiles shown in figure 2 of Gemi\v{s}i\'c Adamov \etal \cite{ADA03} for various cathode materials is satisfactory for the Au cathode.  However, the predicted wings of the H$_\alpha$ lines for their graphite cathode are much larger than observed (not shown here). This reference provides data for another important test of the present model, i.e., the predicted linear dependence of the various components of the Doppler profile on the discharge current is verified by the data of their figure 6. The associated $E/N$ and $J$ range is shown in the present \fref{fig:eonvscurrent}. The papers by this group have argued for the importance of H$_3^+$ ions formed from collisions of H$_2^+$ with H$_2$ in the reaction sequence resulting in H$_\alpha$ excitation.  Because the H$_3^+$ production occurs at $E/N$ and it is destroyed at high $E/N$, the model predicts that only a small fraction of the excitation is by H$_3^+$ collisions with H$_2$, but \fref{fig:kinetics} and figure 1 of \cite{MODEL} show that H$_3^+$ formation and destruction play an important role in the ion, atom, and molecule reaction sequence that leads to H(n=3).  

In summary, the model of the present paper provides quantitative fits to many, but not all, aspects of the H$_\alpha$ emission profiles and spatial distributions measured by Konjevi\'c and coworkers.

\subsection{Experiments of Dexter \etal}\label{subsec:dexter}

The experiments of Dexter \etal \cite{DEX89} test the ability of the model to predict the relative ion fluxes and ion energy distributions for H$^+$, H$_2^+$, and H$_3^+$ reaching the cathode of a low-pressure discharge in H$_2$.  These authors used a mass spectrometer to measure relative ion currents for a glow discharge operating at 2 Torr, 530 V, and a current density of 2.5 A/cm$^{-2}$.  The length of the cathode fall determined from the authors' self-consistent calculations of the electric field is about 6 mm to give an average $E/N$ of 1.3 kTd. This experiment utilizes a ring anode at 40 mm from the cathode. The model uses an effective discharge length of 30 cm.

%Figure 11

The calculated ion-energy distributions are shown by the solid curves of \fref{fig:energydistribution} and the experimental results are shown by points. The authors' Monte Carlo results are indicated by the dotted curves. The model results for H$_2^+$ ions (green curve) and for H$_3^+$ ions (blue curve) at the higher energies show reasonable agreement with relative measurements and with the Monte Carlo calculations of Dexter \etal cite{DEX89}.  At energies below 50 eV, the relatively steep calculated and measured energy distributions for H$_2^+$ ions approach the distribution expected for symmetric charge-transfer collisions \cite{MODEL,RAO95}.  At higher energies, the energy distributions for H$_2^+$ tend to follow those for H$_3^+$ ions, showing the effects of collisional coupling.

The calculated ion-energy distribution for H$^+$ ions (red) is very different than the calculated and experimental results (points) of Dexter \etal.  A similar unexplained discrepancy was reported in \cite{MODEL} with the low current, drift tube measurements of Rao \etal \cite{RAO95} for similar $E/N$ and $pd$ values.  In evaluating these comparisons, it is important to keep in mind the uncertainties associated with this type of measurement, e.g., differences in sensitivity of the detection system found for different ions \cite{MEN06} and differences in attenuation of the ion beams as they leave the exit orifice of the relatively high pressure (2 Torr) discharge chamber and enter the mass spectrometer \cite{HEL74}.  Because of large differences in assumed cross sections, e.g., the authors' questionable assumption of fast two-body conversion of H$^+$ to H$_3^+$, no further attempt is made to compare results for the two models.

\section{High-voltage, hollow-cathode discharges}\label{sec:hollow}

The discharges of interest in this section are perhaps better characterized as ``transparent-cathode'' discharges than hollow-cathode discharges in the text-book sense of electrons oscillating radially in the space charge potential well inside the cathode \cite{DRU40,HOLLOW}. The transition from the conventional hollow-cathode mode to the high-voltage, low-pressure mode for H$_2$ discharges is discussed by Lavrov and Mel'nikov \cite{LAV95,LAV93}.  According to their qualitative observations, as the pressure is reduced the region of highly visible emission moves from the interior of the hollow cathode to the space between the end of the cathode and the anode \cite{AVP10j}. Although the discharge tube is constructed with a ring anode aligned with each end of the hollow cathode, only one of their anodes is electrically connected. The model of this experiment is simplified by making the assumption that at the pressures of interest the tubular cathode can be replaced by a highly transparent planar cathode. This simplification is illustrated in \fref{fig:electrodes}. The H$_\alpha$ radiation produced inside the hollow cathode is neglected. The H$_\alpha$ profiles of figures 2 and 3 of Lavrov and Mel'nikov \cite{LAV95} and figure 9 of \v{S}i\v{s}ovi\'c \etal \cite{SIS05} for low-pressure, hollow cathodes clearly show the high asymmetry resulting from the reduced backscattering expected for a relatively transparent cathode. These comparisons are followed by discussions of somewhat similar discharges in electrostatic confinement devices \cite{KIP08,BOR09,KIP09}.

\subsection{Experiments of Lavrov and Mel'nikov}\label{subsec:lavrov}

The data of Lavrov and Mel'nikov \cite{LAV95,LAV93} are of interest because of the voltage and pressure dependence of their results, their evidence for excitation by at least two species, and the authors proposal that their results show the importance of negative ions.  These authors noted that their Doppler profiles have a relatively weak H$_\alpha$ wing extending to positive wavelength shifts well beyond values expected for H atoms backscattered from H$_2$ molecules. They attribute this observation to excitation of H$_\alpha$ by H$^-$ ions. Consequently, the present model has been extended to include H$^-$ production and loss as described in \sref{sec:kinetics} and \ref{sec:hminusxsect}. \v{S}i\v{s}ovi\'c \etal \cite{SIS05} have also observed similar highly asymmetric H$_\alpha$ profiles and, in addition, have shown the reversal of the asymmetry with the direction of the applied voltage expected for an electric field-dependent excitation and discharge model, such as that of this paper.

%Figure 12

\Fref{fig:lavrov} shows examples of the H$_\alpha$ profiles observed by Lavrov and Mel'nikov \cite{LAV95} observed along the axis of symmetry of their discharge, i.e., along the electric-field lines of the simplified model. The relative magnitudes of the experimental profiles for 0.09 and 0.58 Torr are normalized to current and pressure and then plotted with the same scale factor. The corresponding estimated cathode fall lengths vary from 25 to 4 mm and the cathode to anode separation is taken to be 50 mm.  The calculated curves show good agreement with the experiments covering factors of 3.5 in voltage and 6 in pressure. The data for 0.19 Torr is normalized separately.  As in previous sections, the intensity at negative $\Delta\lambda$ is attributed to excited H(n=3) atoms excited primarily by fast H atoms and fast H$_2$ molecules moving toward the cathode, while the intensities at positive $\Delta\lambda$ are attributed to fast H(n=3) atoms excited by fast H atoms reflected from the cathode.  The relatively small flux of H atoms reflected by the edge of the cathode is treated as a semitransparent cathode effect, i.e, the model assumes that the infinite diameter cathode has a reflected H atom yield of $\approx 50\%$ of that expected for a planar Fe cathode as given in \ref{sec:backscattering}.  The fits of the model to experiment in \fref{fig:lavrov} show the expected need to assume a more diffuse angular distribution as the pressure is increased, i.e., the assumed angular distribution parameter $b$ for approaching fast H atoms is reduced from 10 for 0.09 Torr to 1 for 0.58 Torr.

At the highest pressure of 0.58 Torr the wings of the measured profile in \fref{fig:lavrov}(c) seem to extrapolate smoothly from positive to negative wavelength shifts.  This observation suggests the possibility that multiple scattering and/or large angle scattering events in the excitation chain can cause ions that are initially moving toward the cathode to produce excitation of H(n=3) atoms moving away from the cathode.  Similarly, fast atoms reflected from the cathode may result in H(n=3) atoms moving toward the cathode.  Such events are not included in the present model.

The broken curves of \fref{fig:lavrov} (a) through (c) show that calculated H$_\alpha$ excitation is principally by fast H (solid blue curves) and by fast H$_2$ (dash-dot green curves). The excitation by all ions (dashed purple curves) is significantly smaller.  In \fref{fig:lavrov}(b), the peak in the dash-dot (purple) curves near $\Delta\Lambda=-0.5$ nm is caused by H$_3^+$ and H$_2^+$ ions, while the shelf near -0.7 nm is caused by H$^+$ ions. Thus, the presence of two peaks in the total emission in panel (b) near $\Delta\lambda= -0.6$ nm for 0.19 Torr illustrates the shifting relative importance of excitation by fast H atoms and by fast H$_2$ molecules with pressure. The peak from excitation by ions, especially H$_3^+$ ions, tends to hide the minimum.  Note that because of the large symmetric charge-transfer collision cross section, the H$_2^+$ peak occurs at roughly the same $\Delta\lambda$ (and corresponding velocity) as the peak resulting from excitation by H$_3^+$. As suggested by the calculations of \fref{fig:barbeaufluxes} and the discussion by Dexter \etal \cite{DEX89}, the model shows that most of the H$_2^+$ produced in the low electric-field (negative-glow) region by electron collisions with H$_2$ are converted to H$_3^+$ and that much of this H$_3^+$ is converted back to H$_2^+$ when the ions reach the high-field region.

The proposal by Lavrov and Mel'nikov \cite{LAV95} that production of H$_\alpha$ excitation by H$^-$ ions is significant at positive $\Delta\lambda$ is discussed next. See \ref{sec:hminusxsect} for a summary of relevant cross sections and surface effects. Using a representative surface yield of 0.04 H$^-$ ion per backscattered H atom one obtains the H$_\alpha$ profile at positive $\Delta\lambda$ shown by the dash-double dot (olive) curves in \fref{fig:lavrov}. Because of the very large loss of H$^-$ by collisional detachment at the relatively high pressures of these experiments, this emission is small and is shown after multiplication by factor of 100 \cite{AVP10R}. At 0.19 Torr, the contribution to H$_\alpha$ excitation by H$^-$ produced at the cathode surface is $\sim 70\%$ of the total. The contribution by H$^-$ produced by ion-pair formation in the high field region is $\sim 25\%$ and that by dissociative attachment by electrons (primarily in the low-field, negative-glow region) is $\sim 5\%$. While H$^-$ production by electron capture by reflected fast H atoms leaving the cathode surface can result in H$^-$ with energies up to twice the applied voltage, the model shows little contribution to the H$_\alpha$ profile for energies above that corresponding to the applied voltage. These calculations neglect the potentially important, but unknown, flux of H$^-$ ions emitted by the hollow cathode that would appear with a maximum energy determined by the applied voltage. Thus, the model suggests that H$^-$ induced excitation of H$_\alpha$ has not been observed in these experiments.

\subsection{Experiments of Kipritidis \etal}\label{subsec:kipritidis}

In this section the kinetics model is applied to the Doppler profiles obtained from the low-pressure, hollow-cathode experiments as described by Kipritidis \etal \cite{KIP08,KIP09}. These ``inertial electrostatic confinement'' discharges are designed to build up high densities of fast hydrogen (deuterium) ions and neutral species by trapping the positive ions in a potential minimum created by the hollow cathode. These discharges are assumed to be symmetrical about the center of the hollow cathode, although they sometimes are not. In the model, the hollow cathode is replaced with partially transparent planar cathodes at each end of the actual cathode.  The ring or mesh anodes are replaced with partially transparent planar anodes.  The model does not solve for the discharge behavior inside the cathode, but instead assumes that the ions that strike the surface of the cathode give rise to an electron current that is effectively emitted at the ends of the cathode \cite{ROC84}.  In a simplification suggested by previous studies of these discharges \cite{KIP08,KIP09}, it is assumed that the unknown potential inside the cathode is spatially uniform and at some adjustable fraction of the applied potential \cite{THO97}.  Child's law formulas from \ref{sec:spacecharge} are used to estimate the thickness of the axial cathode sheath adjacent to the ends of the hollow cathode, as illustrated for more conventional discharges in \fref{fig:electricfield}. As in other calculations of this paper, this space charge sheath is replaced by a uniform electric field determined by the effective applied voltage and the sheath thickness. The discharge conditions are similar to those of \sref{subsec:lavrov}, except that the hydrogen pressures are significantly lower and surface effects can become dominant.

The actual calculations assume that the discharge occurs only on the right hand side of the symmetrical electrodes, where the directions are appropriate for \fref{fig:electrodes}.  The asymmetric results are then reflected about the center. In this asymmetric model, the positive ion flux moving leftward toward the cathode builds up from zero at the wall as the result of a nearly uniform electron flux moving to the right toward the anode as expected for a low pressure version of \fref{fig:barbeaufluxes}.  The original ions are mostly H$_2^+$ at near thermal energies and are partially converted to H$_3^+$ ions in the low field region.  Once these ions drift to the high field region of the cathode fall, many of the H$_3^+$ ions are converted to H$_2^+$ and H$^+$, which then produce fast H$_2$ and H. Once past the cathode, the leftward moving ions, but not the fast atoms, turn around in the decelerating electric fields. The assumed (but not verified) significant loss of positive ions to the inside wall of the cathode reduces the effects of positive-ion trapping.  Fast H is produced by particle reflection at the semi-transparent cathodes located at the cathode edges and, especially, at the vacuum-chamber wall.

%Figure 13

\Fref{fig:kipritidis} shows a comparison of calculated and measured Doppler profiles from Kipritidis \etal. The experimental data points in panels (a) and (b) from \cite{KIP08} are for pressures of 23 and 35 mTorr, while the experimental profile in panel (c) is from \cite{KIP09} for 5 mTorr. For the first two pressures, the assumed symmetry of the model is a considerable simplification of the rather asymmetrical anode configuration employed \cite{KIP08}.  The model potential changes from a constant value inside the cathode to a high constant field in the cathode-fall sheath and to a constant value outside the sheath.  From \ref{sec:spacecharge}, the sheath thicknesses are assumed to be 10, 15, and 40  mm for pressures of 35, 23, and 5 mTorr.  The cathode to anode distance and the vacuum chamber radius are assumed to be 100 mm and 240 mm.

The experimental observations in panels (a) and (b) of \fref{fig:kipritidis} were made looking through the ring anode at an angle of 25$^\circ$ with the discharge axis, such that one does not observe radiation from the point of intersection of the electron and particle beams with the wall. The calculated H$_\alpha$ profiles are shown by the solid (red) curves, while the broken curves show various contributions to the total.  The points are sampled from the experimental data.  The agreement between the shapes of the solid curves and experiment is good except for the magnitude of the narrow core, where electron impact excitation dominates. The source of the discrepancies in relative magnitudes of the cores and the wings for these profiles is unknown, but not surprising in view of the use of one-dimensional geometry, etc. In these calculations, the off-axis observations are approximated by adding the axial and transverse contributions calculated using the procedures described in section VA of \cite{MODEL} and applied separately in previous sections of this paper. The calculated axial, transverse, and core contributions are shown by the broken curves of panel (a).  The intensity at $\Delta\lambda > 0.2$ nm is primarily from fast H(n=3) atoms seen by the observer as moving away from the cathode along the discharge axis. The dashed curves of panels (b) and (c) show that the contributions of excitation by fast H atoms and by fast H$_2$ molecules are comparable. The large contribution shown by the dash-double-dot (purple) curve in panel (c) is discussed below. Further tests show that the faster excited H atoms are the result of ion production by electrons between the cathode and the mesh, while nearer the line center the excited atoms are the result of electron induced ionization between the cathode and the ring anode.  Note that because the fast H$^+$, H$_2^+$, and H$_3^+$ ions are created in the potential well near the cathode, they cannot reach the observation region and cause excitation directly.  In the present model of these experiments, the excited atoms decay by emission much too rapidly to move from the high field region or the wall to the observation point. See Appendix B of \cite{MODEL} and \cite{SPATIAL}.

The dashed (green) curve of panel (a) of \fref{fig:kipritidis} shows that the observed H$_\alpha$ for negative wavelength shifts is the result of motion of the H(n=3) atoms perpendicular to the axis of the discharge. This contribution is relatively small for positive shifts. As in the previous models \cite{MODEL,DOPPLER}, this traverse motion is attributed principally to the diffuse angular distribution of H atoms leaving surfaces as the result of bombardment by, in the present case, fast H atoms and H$_2$ molecules.  The model does not calculate the diffuse angular distributions, but uses the adjustable parameter $b$ determining the angular distribution adjusted to best fit experiment \cite{MODEL}, i.e., $b=0.6$ for H(n=3) leaving the cathode and $b=10$ for beam-like H(n=3) approaching the cathode.  The magnitudes of the experimental data in \fref{fig:kipritidis} (a) and (b) are shown with the same scale factor, i.e., they have not been normalized to take into account the expected scaling of roughly factor of four from their products of current times pressure. Thus, the apparent agreement in the relative magnitudes of the wings of these profiles with the model is not understood.

The experimental H$_\alpha$ profile shown by the points in panel (c) of \fref{fig:kipritidis} is for the very low pressure of 5 mTorr and for a very high applied voltage of 30 kV.  It shows observations made near the vacuum wall, looking toward the wall and at an angle of 30$^\circ$ with the discharge axis.  An assumed cathode fall voltage of 12 kV gives the best fit of the model to the experimental profile.  The calculated emission observed at positive values of $\Delta\lambda$ is the result of the excitation of H(n=3) atoms by fast H atoms and H$_2$ molecules approaching the wall, as shown by the dashed (green) and dashed-dot (blue) curves, respectively. Most of these fast neutrals were produced by charge transfer from their analogue positive ions in the region of the potential well. Again, fast positive ions cannot reach the point of observation. Of particular importance for this experiment is the conclusion that the calculated emission at negative $\Delta\lambda$ is primarily the result of H(n=3) atom formation as the backscattered fast H atoms leave the surface of vacuum chamber wall. The process is discussed in \sref{sec:kinetics}. Because the model shows that the excited atoms do not move significantly before radiating, this wall excitation process is generally not observed. The contributions of the axial and transverse components (not shown) are comparable, with the axial component more important at the larger positive frequency shifts and the transverse component at the larger negative shifts. Increasing the cathode fall voltage to 27 kV in the model gives roughly the observed profile width in the far wings, but yields much too flat a profile nearer the line center.

The model is also used to examine the possibility of explaining the far wing ''shelves'' at positive and negative $\Delta\lambda$ beyond 2.2 nm as the result of excitation of H(n=3) by fast H$^-$ ions.  The H$^-$ ions are formed when fast atoms are reflected from the wall, by dissociative attachment, and by ion pair formation. Using estimated cross sections, etc. discussed in \ref{sec:hminusxsect}, the calculated emission is much too small.  Negative ion production inside the hollow cathode is also a possibility. However, only H$^-$ produced as fast H atoms leave the cathode surface can excite H(n=3) atoms with energies higher than that corresponding to the applied voltage.

\subsection{Experiments of Boris \etal}\label{subsec:boris}

In this section the predictions of the kinetics model are compared with the measured energy distributions for D$^-$ ions and with estimated ratios of negative ion and electron fluxes from an electrostatic confinement device described by Boris \etal \cite{BOR09}. The reader is referred to the very extensive published discussions, e.g., the recent results in \cite{BOR09,EMM10}.  Positive ions are injected at the anode and move through a potential with a central minimum, such as shown schematically in \fref{fig:boris}(a).  This potential, to be used in the model below, does not include details such as the potential wells surrounding the individual grid wires. If the injected ions lose small amounts of energy they are trapped in the potential well at relatively high kinetic energies. These trapped positive ions oscillate in the potential well, occasionally producing fast atoms, molecules, or nuclear reactions. Eventually the fast ions collide with the grid wires or undergo charge exchange collisions and drop to low kinetic energies. The low energy ions are collected by the cathode grid wires. The roughly uniform potential distribution inside the cathode grid resulting from ion space charge, is sometimes referred to as a ``virtual anode'' \cite{EMM10}, because it is positive relative to the grid wires. In the present approximation, the equipotential region serves as a ``virtual cathode'' relative to the anode grid.

In the spherically-symmetric device being modelled, the central cathode consists of a highly-transparent grid of 10 cm diam. The anode is a concentric grid of 45 cm diam, centered in a cylindrical Al vacuum chamber of 91 cm diam The geometrical transparency of each grid is assumed to be 0.95. In order to simplify the application of the present kinetics model, the spherically symmetric discharge grids are replaced by partially transparent planar electrodes in the one-dimensional geometry of \sref{sec:survey} and \fref{fig:electrodes}. As shown in \fref{fig:boris}(a), the model assumes a spatially-uniform electric potential inside the grid. This potential plateau is chosen to give the best fit to experiment and is significantly smaller than the applied potential. The electric field immediately outside the cathode grid is approximated by a region of constant electric field strength beginning at the grid for a distance of 6 cm as in \fref{fig:boris}.  This field is only a rough approximation to the space-charge-free potential expected from the considerations of \ref{sec:spacecharge} and \fref{fig:eonvscurrent} and \cite{EMM10}. At larger distances, the electric field is assumed zero on the scale of the energy grid of the calculation. The presumed electric field variations outside the anode grid resulting from biasing the filament and the vacuum wall so as to inject ions into the cathode region are accounted for by assuming that the injected ions drift inward toward the anode grid and that outward moving positive ions are reflected back through the anode and cathode grids. In order to obtain a stable numerical solution to the particle flux equations for the assumed geometry, the assumed transmission of the cathode grid for the lowest energy trapped ions is kept below 0.43 \cite{AVP10x}. The cathode and anode grid transmissions for higher energy ions are assumed equal to their geometrical value of 0.95.  In spite of these rather drastic simplifications, the model serves to illustrate the dominant collision and transport processes leading to H$^-$ production in these extreme conditions. The cross sections and product energy distributions for the H$^-$ formation process are discussed in \ref{sec:hminusxsect}.

%Figure 14

In the following comparison of model results with experiment it is assumed that all aspects of the model developed for hydrogen apply directly to deuterium.  Although this assumption is certain to lead to quantitative errors, there are not expected to be qualitative differences.  \Fref{fig:boris}(b) shows calculated spatial distributions of the sum of the magnitudes of the left- and right-directed particle fluxes for the case of 2 mTorr pressure and 70 kV applied voltage presented in figure 4 of \cite{BOR09}. These calculated fluxes are normalized to unit positive ion flux entering at low energies at the anode grids.  The large normalized fluxes calculated for H$_2^+$ and H$^+$ are misleading because more than 90\% of these ion fluxes are in the lowest energy bin of the model and do not have sufficient energy to form H$^-$, etc. A better measure of the effectiveness of the ion trapping is to note that the calculated H$_2^+$ and H$^+$ fluxes inside the cathode and having energies above the lowest energy bin (1500 eV in our numerical scheme) are each about equal ($\pm 20\%$) to the total ion flux injected at the anode.

The results plotted in \fref{fig:boris}(b) show that the dominant ion is H$_2^+$ and the dominant fast neutral is the H$_2$ molecule. The calculated results are insensitive to the proximity of the assumed value of the grid transmission for low energy ions to the value causing numerical instability. The electron flux is calculated from the ion fluxes striking the cathode grid \cite{KRU08} using the assumed grid transparency and the electron yield per H$^+$ ion for Mo \cite{THO85}, which is chemically similar to the W (but not Re) used experimentally. This figure also shows that the calculated ratio of the H$^-$ flux to the electron flux is $\sim 0.3\%$ compared to the experimentally estimated value of $\sim 3\%$ \cite{BOR09}. This result is very insensitive to the assumed grid transmission. For the present parameters, the discharge is sustained by the external ion source. Heavy particle ionization makes an important contribution to the ion production, but is not sufficient to balance losses as assumed by Emmert \etal \cite{EMM10}

The principle observable in this experiment is the energy distribution of H$^-$ ions as measured by an energy analyzer at the vacuum chamber wall. The points of \fref{fig:boris}(c) show the measured energy distribution for H$^-$ ions for an applied voltage of 70 kV at a pressure of 2 mTorr.  Boris \etal \cite{BOR09} attribute the peaks in their measured energy distribution near 40 and 50 keV to H$^-$ formed in collisions of H$_3^+$ and H$^+$ with H$_2$, respectively. The calculated energy distribution gives a series of very sharp peaks superimposed on a background. For purposes of comparison with experiment, the calculated curves have been folded into a Gaussian, such as might arise from local spatial variations in the electric potential. The energy analyzer is assumed to effectively sample the H$^-$ flux reaching the anode grid. The model makes no allowance for possible spatial or energy variations in the H$^-$ detection efficiency in the experiment. The dashed (red) curve of \fref{fig:boris}(b) shows that most of the H$^-$ is formed in collisions of fast H atoms with H$_2$. The dotted (brown) curve shows that there is a significant contribution to the $\approx 40$ keV peak from particle reflection at the surfaces of cathode grid wires. The dash-dot (olive) curve shows that H$^-$ production in H$^+$ + H$_2$ collisions makes a significant contribution to the peak near 59 keV.  Production of H$^-$ in fast H$_2$ and H$_2^+$ collisions with H$_2$ are smaller and production by H$_3^+$ is too small to plot.

The energy scale of the calculated peaks in \fref{fig:boris}(c) is set by assuming the effective plasma potential plateau inside the cathode grid to be $\approx 40$ kV negative relative to the anode grid wires and vacuum wall. The fit is much worse if the potential plateau is assumed to be below the anode potential by the applied 70 kV. The decrease in H$^-$ flux at energies above the high energy peak at $\approx 58$ keV is in qualitative agreement with experiment. Calculations show that the $\approx 58$ keV peak increases in magnitude relative to the $\approx 40$ keV peak as the applied voltage increases, as is observed in the experiments \cite{BOR09}. The model also shows a decrease in relative magnitude of this higher energy peak with increasing pressure as shown in figure 7 of \cite{BOR09}. However, the model fails to predict the observed ratio of peak heights for the $\approx 40$ and $\approx 58$ keV peaks.

Prominent features of the calculated H$^-$ energy distributions are the low energy peaks, e.g., those near 15 keV in \fref{fig:boris}(c). If the energy analyzer is assumed to effectively sample the H$^-$ beam at the outer edge of the high field region rather than at the anode as assumed in \fref{fig:boris}(c), the H$^-$ flux at energies below $\approx 40$ keV is reduced by more than an order of magnitude. This change occurs because the low energy peaks are the result of H$^-$ formation principally by fast H atoms in the space between the high field region and the vacuum wall.  Note that this region is calculated to be free of energetic positive ions. The model does not reproduce the very small relative values of H$^-$ flux found at the lowest energies reported experimentally \cite{BOR09}, i.e., just below the $\approx 40$ keV peak.

A much more thorough analysis of the model and experiments is necessary to ensure the applicability of the model. For example, the calculated energies and relative magnitudes of the peaks in the H$^-$ energy distribution are systematically shifted relative to experiment. No attempt has been made to estimate the role of nonlinear processes, such as charged particle recombination or nuclear reactions involving collisions between energetic particles \cite{EMM10}. Spatial and spectral scans of the H$_\alpha$ emission should help define some of the geometrical and electrical parameters, as shown by the analyses in \sref{subsec:barbeau}, \sref{subsec:cvet}, and \sref{subsec:kipritidis}.  As shown throughout this paper, the wings of the H$_\alpha$ profiles are good diagnostics for high-energy hydrogen particles.

\section{Other experiments}\label{sec:other}

Babkina \etal \cite{BAB05} determined the energy spectrum and enhanced far-wing of H$_\alpha$ emission of fast H atoms produced by various hydrogen ions when backscattered from a stainless steel surface biased negatively with respect to a microwave plasma source.  Their H$_\alpha$ profiles also provided evidence of H$^-$ formation from energetic hydrogen reflected from their negative electrode.  The present model does not apply to their experiment because it does not have a sufficiently developed data base for collisions in H$_2$-Ar gas mixtures.

The shape of the H$_\alpha$ profile seen with an optical probe at large distances from the cathode by Bharathi \etal \cite{BHA09} is similar to the axial profiles of Barbeau and Jolly \cite{BAR90} at about the same pressure and discharge voltage. See \sref{subsec:barbeau}. Bharathi \etal discuss the various reactions and excitation processes, but do not calculate H$_\alpha$ profiles.  Because of expected strong departures of their discharge from one dimensional geometry when the optical probe is moved to their ``near cathode'' position, no attempt has been made to model their H$_\alpha$ profiles.

Another type of dc glow discharge in hydrogen is the ``hollow cathode'' with an internal anode examined by the mass spectrometer studies of M\'endez \etal \cite{MEN06}.  These authors analyze their results in terms of ion-molecule reactions that take place at energies determined by the wall temperature. They argue that one can neglect reactions of energetic hydrogen species in the cathode sheath or in the effusive gas flow region. No attempt has been made to adapt the present model to these experiments.

Next, consider pure hydrogen discharges in which ions are drawn from a plasma, accelerated, strike a planar cathode, and may be sampled by a mass spectrometer. Some examples are: Heim and Stori \cite{HEI92}, Hallil \etal \cite{HAL00}, Gans \etal \cite{GAN02}, Babkina \etal \cite{BAB05}, and Schiesko \etal \cite{SCH09}. In general, one expects the average energies of the ions in the discharge source to be a few eV as determined by the ambipolar field generated by the electrons. These discharges produce H$_2^+$ by electron impact and, except at very low pressures, the low energy H$_2^+$ are rapidly converted to H$_3^+$. The kinetics model of the present paper is particularly appropriate for predicting the further reactions that occur in the space charge sheath, but no comparisons have been attempted.

The few dc experiments of Mills \etal \cite{MIL09} and Phillips \etal \cite{PHI09} for pure H$_2$ are not useful for quantitative testing of models of the source of fast H(n=3) atoms emitting far-wing H$_\alpha$ radiation.  For example, the H$_\alpha$ profile of figure 14 of \cite{MIL09B} was obtained at an unknown pressure and unknown position relative to the beginning of the negative glow. The best one can say is that this profile is qualitatively similar to the transverse profiles of \fref{fig:barbeauprofiles}(a) and \fref{fig:cvetprofiles}(b). This author does not know of any discrepancies between the predictions of the present model and their qualitative experimental results for dc discharges in hydrogen. It is expected that an extension of an electric-field-based model will explain well-characterized measurements of H$_\alpha$ profiles in their rf discharge geometries, e.g., figure 9 of \cite{PHI07}.

Other hydrogen plasmas to which the present kinetics model is expected to be applicable are the aurora observed in the hydrogen-rich outer planets \cite{REG94} and controlled fusion plasmas, especially as one approaches the walls \cite{HEY04}. However, no examples have been analyzed that illustrate this applicability because of the difficulty in constructing simple models of the very complex geometries, the spatial and temporal fluctuations of the electric and magnetic fields, and the variations in gas composition and densities.

\section{Discussion}
\label{sec:discuss}

The comparisons of model predictions with experiment in this paper have demonstrated the usefulness of a simplified model of the kinetics of energetic hydrogen ions, atoms, and molecules for quantitatively explaining observations of H$_\alpha$ emission from low-pressure H$_2$ discharges ranging from short-gap, parallel-plane discharges; through glow discharges of various geometrical complexities; to beam-like, electrostatic-confinement discharges. The comparisons have emphasized the shapes of H$_\alpha$ Doppler broadened profiles, relative spatial distributions of spectrally integrated emission, and the energy distributions of mass-identified ion fluxes at the cathode.  Similarly successful comparisons of predicted and measured absolute emission at low current densities were made earlier \cite{MODEL,SPATIAL,DOPPLER}.

The comparisons of calculated and observed H$_\alpha$ profiles and spatial distributions show the importance of H(n=3) excitation in collisions of fast H and fast H$_2$ with H$_2$ for these discharges.  The calculated excitation by H$^+$, H$_2^+$, and H$_3^+$ ions is generally less important, although the model results for ions are less certain because of the absence of low energy excitation cross sections.  The predictions of the role of H$^-$ ions show that, although the H$^-$ can lead to H(n=3) with energies higher than expected from the applied voltage, the experiments analyzed show no convincing evidence for such fast excited atoms. The direct demonstrations of linearity of the magnitude of the H$_\alpha$ emission at low H$_2$ pressures over current density segments of up to four orders of magnitude and the applicability of a model linear in current density over eight orders of magnitude in current density is an important result of this paper. The test is consistent with the basic assumption of the model that the kinetics of these discharges includes a sequence of binary collisions of active species, i.e., ions, atoms and molecules in ground or excited states, with the undissociated hydrogen gas. This test shows that collisions between two or more particles created by the discharge are of little importance. Therefore, this result should put to rest claims by Mills, Phillips, and coworkers \cite{MIL09,PHI09} that the ``excessively broadened'' H$_\alpha$ emission from their dc discharges is a result of energy made available by collision of a dissociation product, e.g., an H atom and another excited atom or dissociation product.

The cross-section set and surface-interaction probabilities for hydrogenic species have been extended to include the production and loss of H$^-$ and more accurate representations of published backscattered H atom, H$^-$ ion, and H(n=3) atom fluxes. Excitation of H$_\alpha$ and negative ion formation probabilities at surfaces are very difficult to characterize and are currently based on very limited data.  The simplified model of this paper still treats the angular distributions of the H(n=3) atoms as an adjustable parameter.  The differential scattering data needed to overcome this deficiency includes extensive sets of differential cross sections for the numerous processes considered so as to improve on published models using very much simplified sets \cite{PET05}. Cross sections for excitation of the Balmer series and UV in H$_x^+$, H, H$^-$, and H$_2$ collisions with H$_2$ (and H atoms) at energies below 2 keV are particularly important and, except for H + H$_2$, are currently only educated guesses. Total rates of energy loss at energies below 10 keV by these species in H$_2$ are poorly known.

The different observable quantities discussed in this paper provide a range of approaches to learning about the transport and reactions of the hydrogenic particles in hydrogen discharges and plasmas.  The H$_\alpha$ Doppler profiles are a measure of the velocity distributions of not only the excited atoms, but of the ions, atoms, and molecules that produced them. The spatial distributions of H$_\alpha$ emission have proved to be a sensitive technique for demonstrating the importance of processes such as the reflection of fast H atoms from surfaces. Positive ion energy distributions are expected to be useful measures of the dominate process for energy gain by heavy particle from the electric field, although there are presently serious discrepancies between the model and experiment. The negative ion energy distributions are a potentially valuable diagnostic for probing the electric field distribution in various hydrogen discharges. From the atomic physics point of view, the wings of transverse H$_\alpha$ profiles provide a measure of the angular distribution of H(n=3) atoms, e.g., near a surface where the angular distribution of H(n=3) atoms is usually dominated by diffuse atom emission. The low electron densities near the cathode result in a increasing contribution of wings relative to core in the negative glow and offer the possibility of study of the small wavelength shifts caused by heavy particles during target species excitation.

It is important to the future application of the kinetics model of this paper that sets of differential cross section be developed for the dominant processes.  These processes include elastic scattering, including symmetric charge transfer; inelastic processes, such as vibrational excitation and Lyman series excitation and ionization; and ion molecule reactions, including ion and proton transfer.  Although the present hydrogen kinetics model has been applied only to H$_\alpha$ production in the various moderate current discharges of this paper, the kinetics model is readily extended to the production of Lyman lines and the near UV continuum.

\begin{ack}

The author wishes to express his thanks to A. Gallagher for continued helpful discussions and for a very thorough critique of the manuscript.  I wish to thank N. Konjevi\'c for supplying tables of H$_\alpha$ line profile data and for helpful correspondence, J. Jolly for high quality drawings of profiles and discharge voltages, and J. Kipritidis for tables of emission profiles. The preparation of this paper was supported in part by JILA.

\end{ack}

\appendix
\section{H$^-$ properties} \label{sec:hminusxsect}

%Figure 15

\Fref{fig:negativeionqs}(a) shows recommended cross sections for collisions of H$^-$ with H$_2$.  The cross sections for momentum-transfer collisions and for electron detachment are the same as in \cite{PHE90}, except that these and other cross sections have been extended to 100 keV using data from \cite{HUN90}. Of particular interest is the cross section for the production of H(n=3) in collisions of H$^-$ with H$_2$ shown by the dashed curve of \fref{fig:negativeionqs}(a).  This cross section is based on the measurements of Geddes \etal \cite{GED81} at energies from 5 to 25 keV. At the very important lower energies, it is scaled from the proposed cross section for H$_\alpha$ excitation by H$^+$ from \cite{MODEL}.  This long extrapolation leads to considerable uncertainty in the predicted H$_\alpha$ excitation by H$^-$ in \sref{subsec:lavrov}.  Obviously, there is a need for direct measurements of this process at lower energies.

\Fref{fig:negativeionqs}(b) shows cross sections for H$^-$ production used in the model of H$_2$ discharges in this paper. The projectiles and the associated references utilized are H \cite{HUN90,ZYL81,BRE95}, H$_2$ \cite{HAY88}, H$^+$ \cite{HUN90,WIL66a,SAL10}, H$_2^+$ \cite{WIL66,ALV76}, and H$_3^+$ \cite{WIL66,ALV76}. Note that the cross sections for H$^-$ formation in collisions of H atoms with H$_2$ from Van Zyl \etal \cite{ZYL81} are significantly larger, particularly at low energies, than those for H$^-$ formation in collisions of H$^+$ \cite{HUN90} and H$_2$ \cite{HAY88} with H$_2$. The present model interprets the literature as showing that for all projectiles the H$^-$ retains very nearly the incident projectile velocity. These high velocities for the product H$^-$ are particularly important for modelling the experiments of Boris \etal \cite{BOR09}.  In several beam type experiments, a two-step process leading to H$^-$ is observed \cite{DUR72,FOU72,LEE87}. The present model assumes that fast H atoms are the intermediate species, rather than excited states.

\section{Space charge effects} \label{sec:spacecharge}

For the relatively high current and charged-particle densities of the cathode fall, it is necessary to know or estimate space-charge electric fields as described by Poisson's equation. This equation for the voltage $V(\rho)$ normalized to the cathode fall voltage $V_c$ can be written as
\begin{equation}
{\frac{d^2 v(z)}{d z^2}} = -{\frac{J_t}{V_c \epsilon_0}}
\left[{\frac{j_i(z)} {w_i(E/n)}- {\frac{j_e(z)}{w_e(E/n)}}} \right],
\label{eq:poisson}
\end{equation}
where $v(z) = V(z)/V_c$, $w_i(E/N)$ and $w_e(E/N)$ are the ion and electron drift velocities, and $z$ is the distance from the cathode.  Here $J_t$ is the total current density and $\epsilon_0$ is the permittivity of free space.  In the vicinity of the cathode, the electron term in \eref{eq:poisson} can usually be neglected. The drift velocities are assumed to be determined by the local $E/n$ ratio given by
\begin{equation}
\frac{E}{N} = - \frac{V_c} {N} \frac{d v(z)} {d z}. \label{eq:field}
\end{equation}
Following von Engle and Steenbeck \cite{ENG34}, the assumption is made that the electric field varies linearly over a distance equal to the thickness of the cathode fall $d_c$, i.e., $E(z)= 2(V_d/d_c)(1-z/d_c)$.  Finally, it is assumed that the total current and the ion current are related by ${J_t}/{j_i} = 1 + \gamma_{eff}$, where $\gamma_{eff}$ is effective value of the electron yield per ion arriving at the cathode \cite{DRU40,RAY79}. The solution to \eref{eq:poisson} can now be expressed as
\begin{equation}
{J_t}/{p^2} = 2 \epsilon_0 {V_d w_i (1+ \gamma_{eff})}/{(p d_c)^2},
\label{eq:currentdensity}
\end{equation}
where the results are expressed as a function of pressure $p$ instead of gas density. The experimental points shown in \fref{fig:electricfield} demonstrate that the assumption of a linear decrease of the electric field with position is valid in some cases, but not in others.  In the absence of better simple models, the linear assumption is used and one expects a significant uncertainty in the predictions.

The dashed line of \fref{fig:eonvscurrent} marking the onset of space charge distortion is calculated using \eref{eq:currentdensity} assuming a $10\%$ decrease in field at the anode at $d$ from the cathode. The solid line marking a fully developed cathode fall assumes the field decreases to zero at the end of the cathode fall at the cathode fall thickness $d_c$. The $E/N$ at the cathode is found to be proportional to $(J/N)^{4/3}$ and to $(N d_c)^{m}$, where $J$ is the current density. The exponent $m$ changes from $1/3$ to $2/3$ as the ion motion changes from free fall to constant mobility. The solid line is shown for a representative gas pressure of 0.3 Torr and $p d_c = 0.2$ Torr cm as found experimentally \cite{FRA56}, while the dashed line is for $p = 0.3$ Torr and $d=4$ cm as are typical in low current experiments \cite{MODEL,SPATIAL,DOPPLER}.

Using \eref{eq:currentdensity} and the electric field at the cathode and current density data of Ganguly and Garscadden \cite{GAN94} shown in \fref{fig:electricfield}, the positive ions have a mean velocity of $3.7\times10^5$ m/s near the cathode. The dominant positive ion is calculated to be H$_2^+$. This velocity is to be compared with $1.4\times10^5$ m/s calculated from the mobility for 100 eV H$_2^+$ in H$_2$ \cite{PHE90} and $5.8\times10^5$ m/s calculated for free-fall motion of H$_2^+$ through the cathode fall voltage.

%
\section{Backscattering of fast H atoms} \label{sec:backscattering}

This Appendix presents fits to published fast H atom energy distributions and backscattered fractions obtained using Monte Carlo techniques \cite{ECK84}.  The empirical analytic fit to the set of calculated energy distributions is
\begin{eqnarray}\label{eq:backscatter}
F = 0.52&&\epsilon_r^{0.4}
   [1 + 900 \epsilon_r^{(17.9/\epsilon_0^{0.3})}(30 + \epsilon_0)^{-1}]
   [1 + \epsilon_0^2 \epsilon_r^{1.5}/12500000]^{-1} \\ \nonumber &&(1 - \epsilon_r/0.95)(1 - \epsilon_r/1.05)^{-1}
\end{eqnarray}
for $\epsilon_r < 0.95$.
$F=0$ for $0.95<\epsilon_r <1$. Here $\epsilon_r$ is the energy of the backscattered H atom relative to the energy of the incident H$^+$ ion $\epsilon_0$.

%Figure 16

The predictions of this expression (smooth curves) for H$^+$ in Ni show agreement with Monte Carlo calculations for D$^+$ incident on Ni shown by the points in \fref{fig:backscatter}(a).  Note the close equality of backscattering expected for H$^+$ and D$^+$ \cite{ECK84}.  The empirical expression also agrees well with the Monte Carlo predictions of Oen and Robinson \cite{OEN76}. The agreement with the experiments of Aratari and Eckstein \cite{ARA89} is only fair. As a check on the empirical expression, the numerical integrations of \eref{eq:backscatter} are compared with particle ``particle reflection coefficient'' R$_N$ data in \fref{fig:backscatter}(b).  There is good agreement with the empirical fit to published R$_N$ data discussed in \cite{MODEL}.

\noappendix

%figure 1
\newpage
\begin{figure}[p]
\includegraphics[width=76mm]{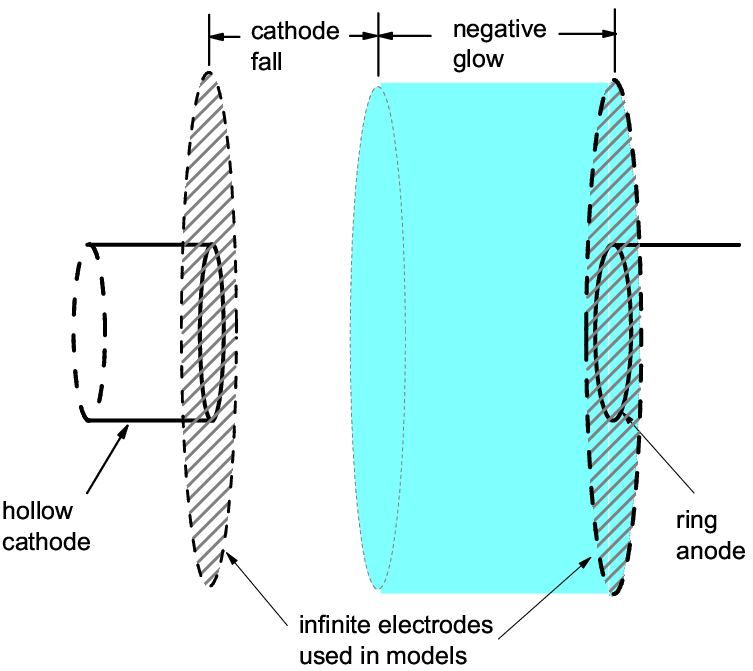}
\end{figure}

\begin{figure}
\caption{\label{fig:electrodes} Schematic of low-pressure glow-discharge experiments analyzed in this paper, showing various electrode configurations discussed in the text.}
\end{figure}
\clearpage

%Figure 2
\newpage
\begin{figure}[p]
\includegraphics[width=76mm]{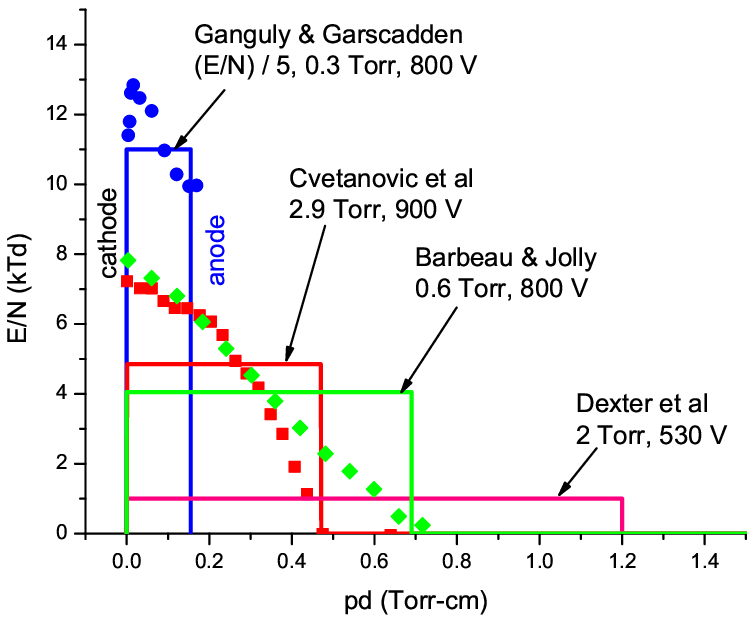}
\end{figure}

\begin{figure}
\caption{\label{fig:electricfield} Measured (points) and approximate (lines) spatial distribution of electric field for Ganguly and Garscadden \cite{GAN91} - circles (blue), for Cvetanovi\'c \etal \cite{CVE05} - squares (red), Barbeau and Jolly \cite{BAR90} - diamonds (green) , and Dexter \etal \cite{DEX89}.  For the last three cases, the extent of the high-field region shown is equal to the stated cathode-fall thickness and the anode locations are well beyond the cathode fall.}
\end{figure}
\clearpage

%Figure 3
\newpage
\begin{figure}[p]
\includegraphics[width=76mm]{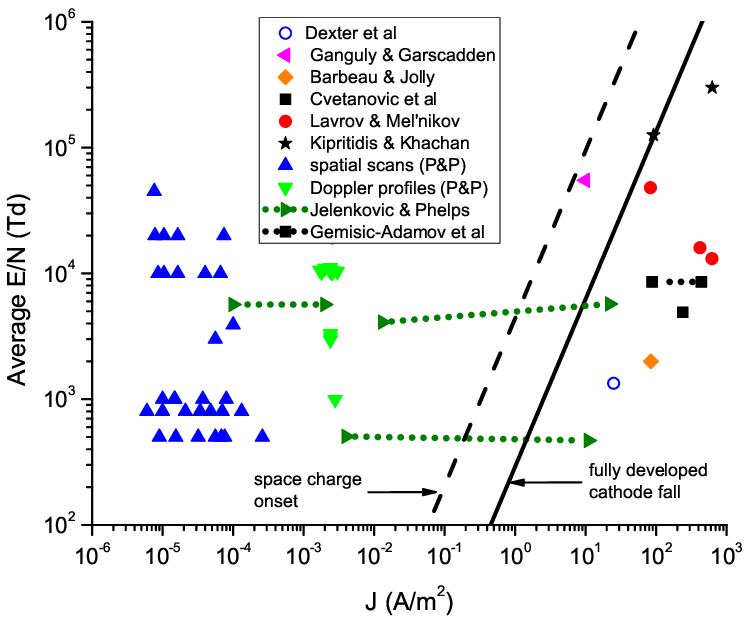}
\end{figure}

\begin{figure}
\caption{\label{fig:eonvscurrent} Average E/N in cathode region versus discharge-current density for low-pressure hydrogen dc discharges considered in this paper. The dashed line indicates the presence of significant (10\%) space-charge distortion of the electric field, while the solid line shows the predicted transition to a fully developed cathode fall. These curves are for typical values of pressure times electrode spacing. The dotted lines indicate direct tests of the linear variation of the H$_\alpha$ excitation with discharge current density.}
\end{figure}
\clearpage

%Figure 4
\newpage
\begin{figure}[p]
\includegraphics[width=76mm]{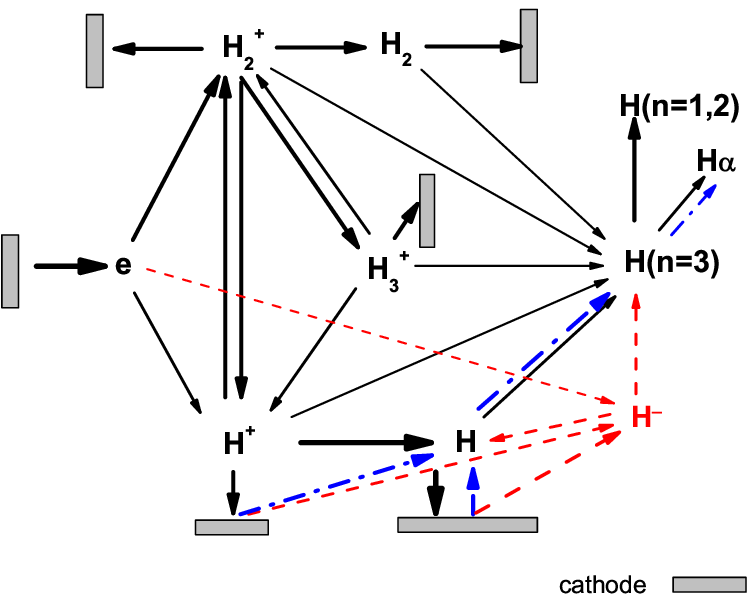}
\end{figure}

\begin{figure}
\caption{\label{fig:kinetics} Schematic of reactions of the indicated species with H$_2$. The solid (black) and dot-dash (blue) lines show reaction paths included in the earlier model \cite{MODEL} for species approaching the cathode and leaving the cathode, respectively. The dashed (red) lines show reactions added when negative ions are included in the model. Heavy-particle induced ionization, energy loss, and excitation of H(n=3) at surfaces are not shown. The relative rates of the various processes are roughly proportional to the widths of the lines.}
\end{figure}
\clearpage

%Figure 5
\newpage
\begin{figure}[p]
\includegraphics[width=76mm]{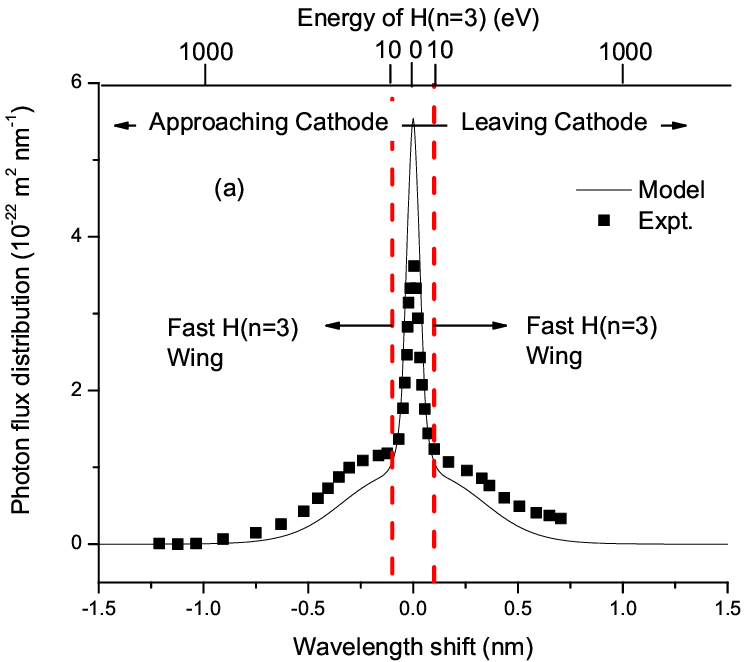}
\end{figure}

\begin{figure}[p]
\includegraphics[width=76mm]{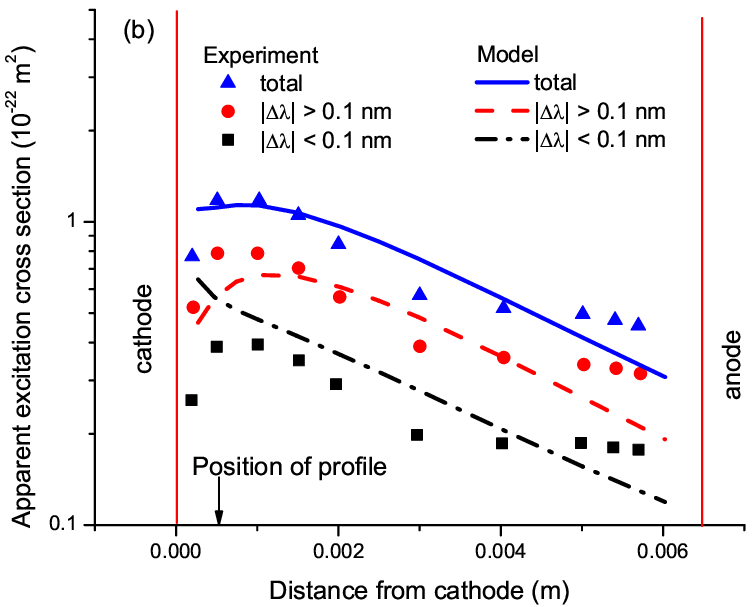}
\end{figure}

\begin{figure}
\caption{\label{fig:ganguly} (a) Wavelength dependence of H$_\alpha$ emission transverse to the discharge axis for a position 0.5 mm from the cathode. The vertical dashed lines show the authors' transition from the core to the wings of the profile. (b) Spatial dependence of H$_\alpha$ emission from obstructed discharge in hydrogen.  The points are from the experiments of Ganguly and Garscadden \cite{GAN91}. The curves are from the model.}
\end{figure}
\clearpage

%Figure 6
\newpage
\begin{figure}[p]
\includegraphics[width=76mm]{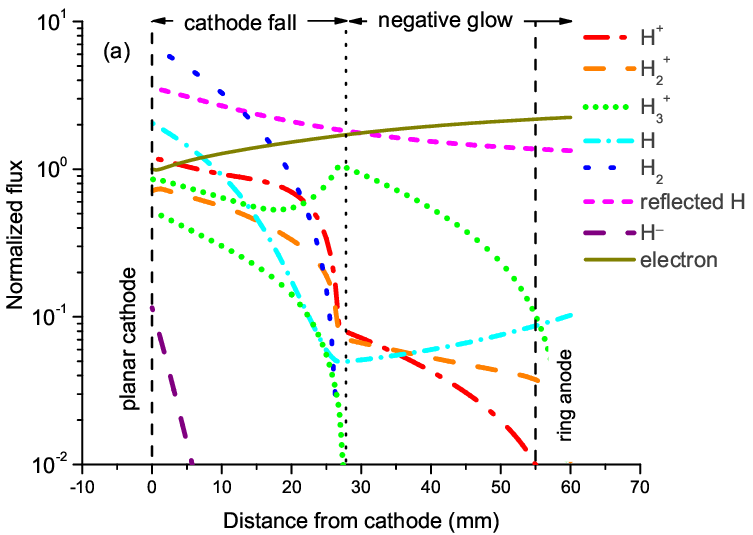}
\end{figure}

\begin{figure}
\caption{\label{fig:barbeaufluxes} Calculated ion, atom, and molecule fluxes for conditions of experiment by Barbeau and Jolly \cite{BAR90} at 1100 V and 0.27 Torr.}
\end{figure}
\clearpage

%Figure 7
\newpage
\begin{figure}[p]
\includegraphics[width=76mm]{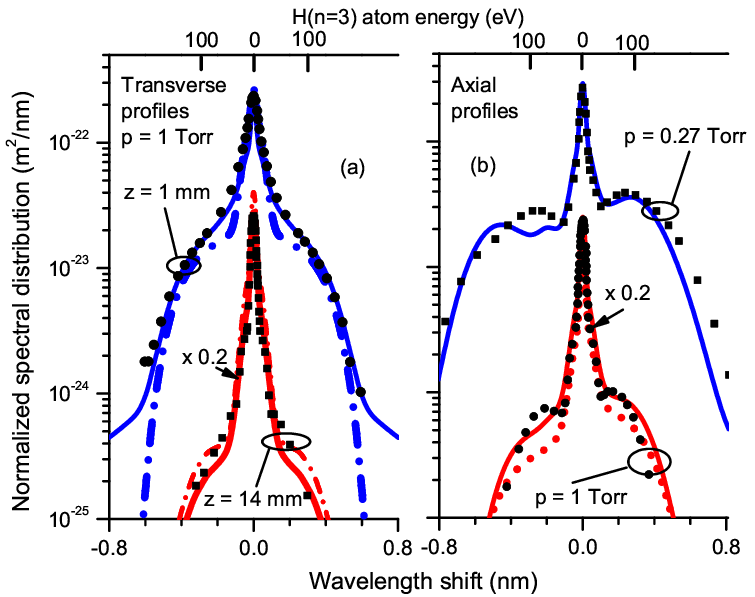}
\end{figure}

\begin{figure}
\caption{\label{fig:barbeauprofiles} Comparison of calculated spectral distribution of H$_\alpha$ emission (curves) with observations (points) by Barbeau and Jolly \cite{BAR90}.  (a) The profiles measured transverse to the electric field at positions of 1 and 14 mm from the cathode for 1 Torr are adjusted in magnitude to the calculated curves with the same scale factor. The solid curves and dotted curves are calculated using the dispersion and Gaussian approximations for dissociative excitation of H(n=3), respectively. (b) Axially observed points for pressures of 1 and 0.27 Torr and a current of 3 mA.}
\end{figure}
\clearpage

%Figure 8
\newpage
\begin{figure}[p]
\includegraphics[width=76mm]{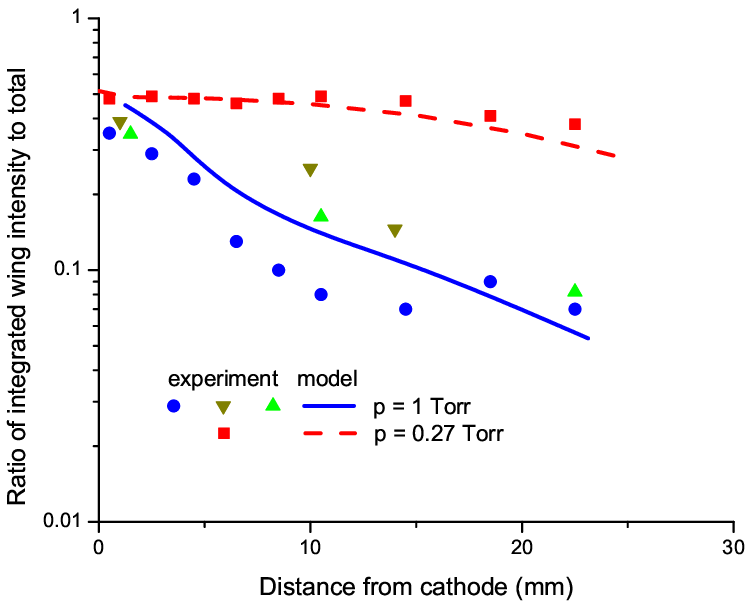}
\end{figure}

\begin{figure}
\caption{\label{fig:barbeaufraction} Measured experimental points from Barbeau and Jolly \cite{BAR90} and calculated curves of the spatial distributions of the fraction of the H$_\alpha$ emission at wavelength shifts greater than 0.1 nm.}
\end{figure}
\clearpage

%figure 9
\newpage
\begin{figure}[p]
\includegraphics[width=76mm]{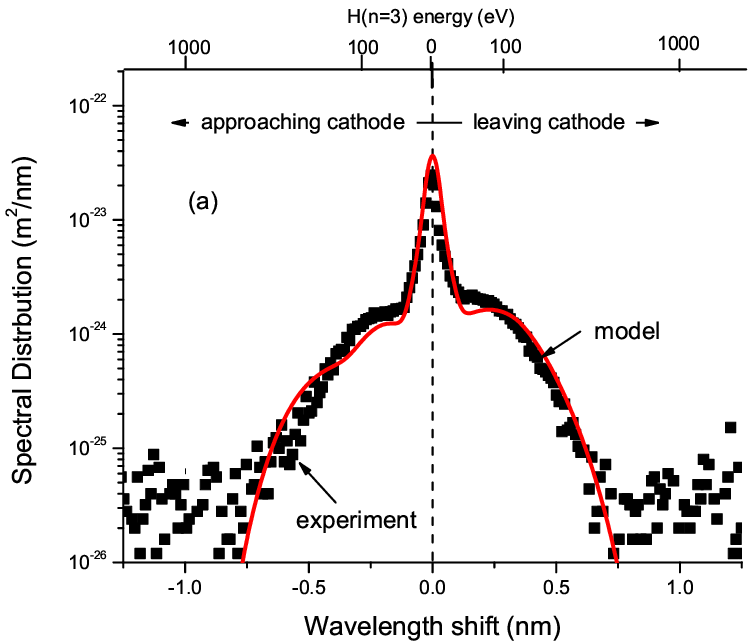}
\end{figure}

\begin{figure}[p]
\includegraphics[width=76mm]{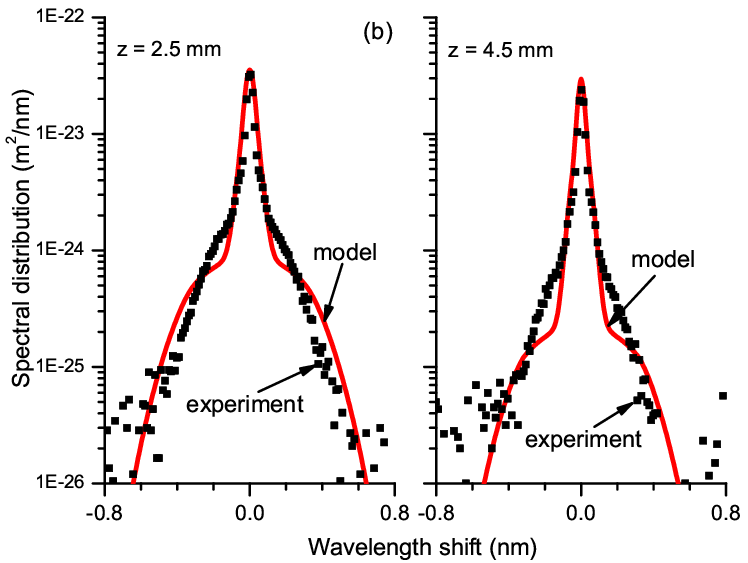}
\end{figure}

\begin{figure}
\caption{\label{fig:cvetprofiles} (a) Comparison of calculated and measured Doppler profiles observed parallel to the electric field. (b) Comparison of calculated and measured Doppler profiles observed transverse to cathode normal from two positions in the negative glow region.  These data are from Cvetanovi\'c \etal \cite{CVE05}.}
\end{figure}
\clearpage

%Figure 10
\newpage
\begin{figure}[p]
\includegraphics[width=76mm]{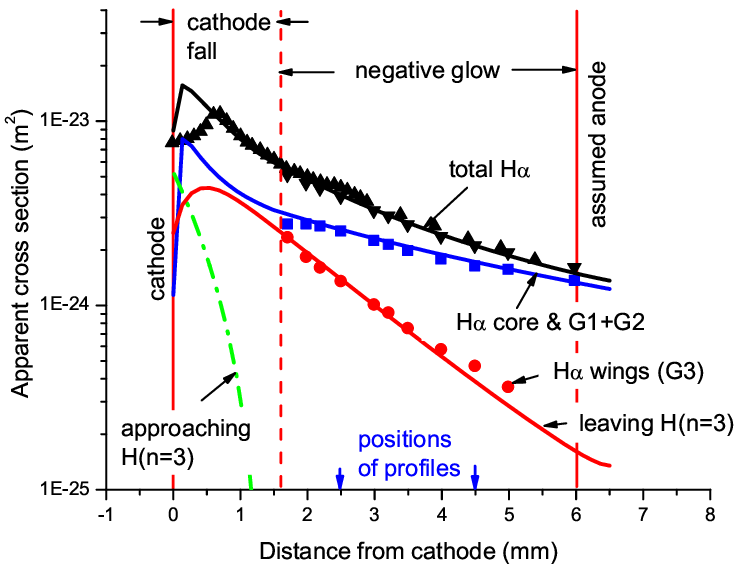}
\end{figure}

\begin{figure}
\caption{\label{fig:cvetspatial} Measured and calculated spatial distribution of H$_\alpha$ emission. The points are relative values from the experiment of Cvetanovi\'c \etal \cite{CVE05}.  The smooth curves are calculations.}
\end{figure}
\clearpage

%Figure 11
\newpage
\begin{figure}[p]
\includegraphics[width=76mm]{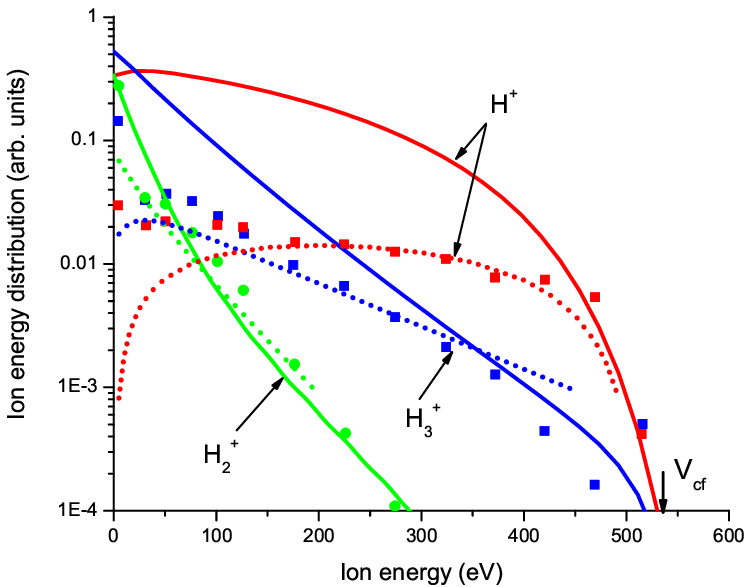}
\end{figure}

\begin{figure}
\caption{\label{fig:energydistribution} Measured and calculated ion-energy distributions at the cathode for $p = 2$ Torr and $V = 530$ V from the experiment of Dexter \etal \cite{DEX89}.  The points are from their experiment and the dotted curves are smoothed fits to their Monte Carlo results. The smooth curves are predictions of the present model.}
\end{figure}
\clearpage

%Figure 12
\newpage
\begin{figure}[p]
\includegraphics[width=76mm]{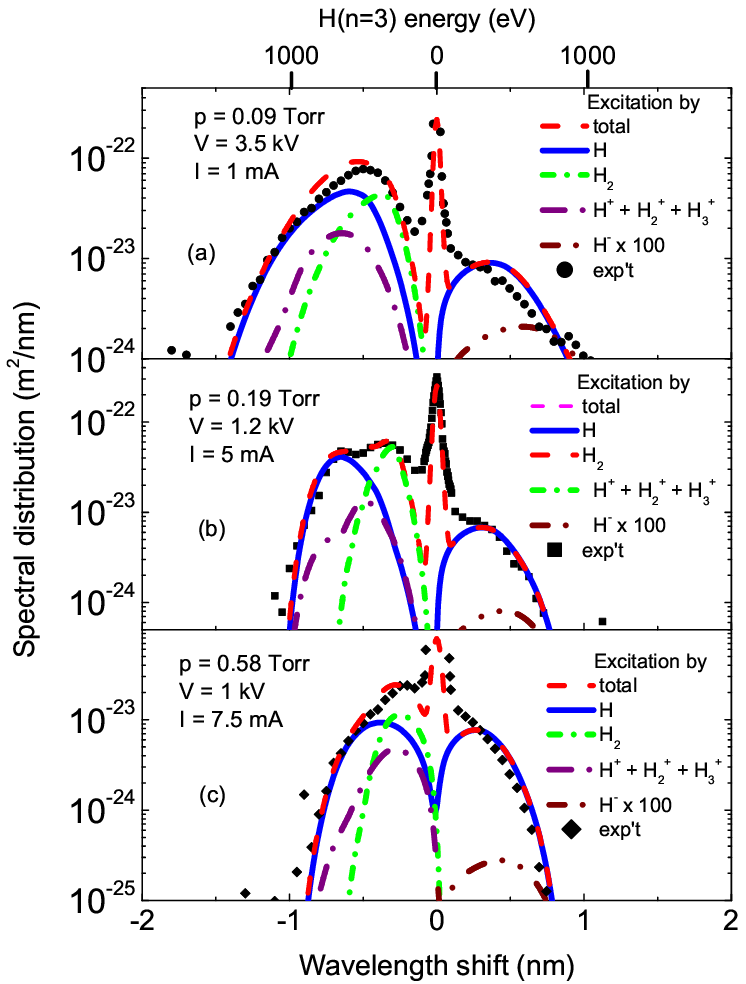}
\end{figure}

\begin{figure}
\caption{\label{fig:lavrov} Comparison of calculated (curves) and measured (points) axial Doppler profiles from Lavrov and Mel'nikov \cite{LAV95} for various pressures, voltages, and currents.  The curves also show the contributions of excitation by various species to the production of H(n=3) atoms.}
\end{figure}
\clearpage

%figure 13
\newpage
\begin{figure}[p]
\includegraphics[width=76mm]{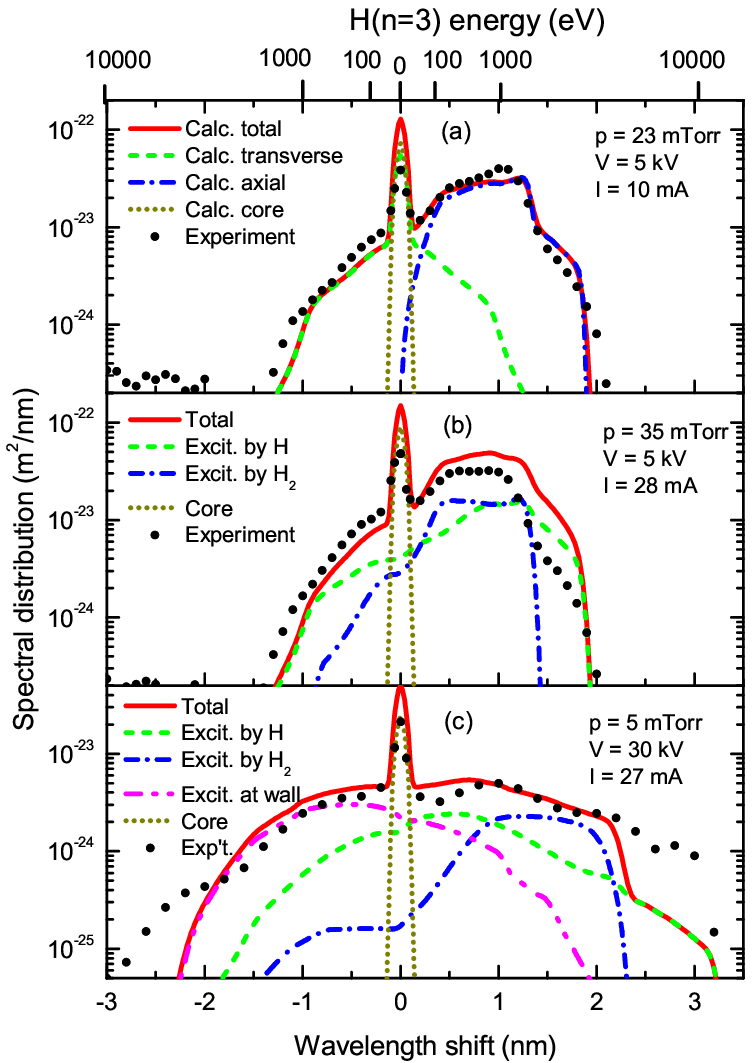}
\end{figure}

\begin{figure}
\caption{\label{fig:kipritidis} Comparison of calculated and measured axial Doppler profiles from Kipritidis \etal. The experimental points of panels (a) and (b) are for the geometry of figure 3 of \cite{KIP08} and those of panel (c) are from figure 6 of \cite{KIP09} for the geometry of the associated figure 3. The smooth curves show calculations discussed in the text.}
\end{figure}
\clearpage

%figure 14
\newpage
\begin{figure}[p]
\includegraphics[width=76mm]{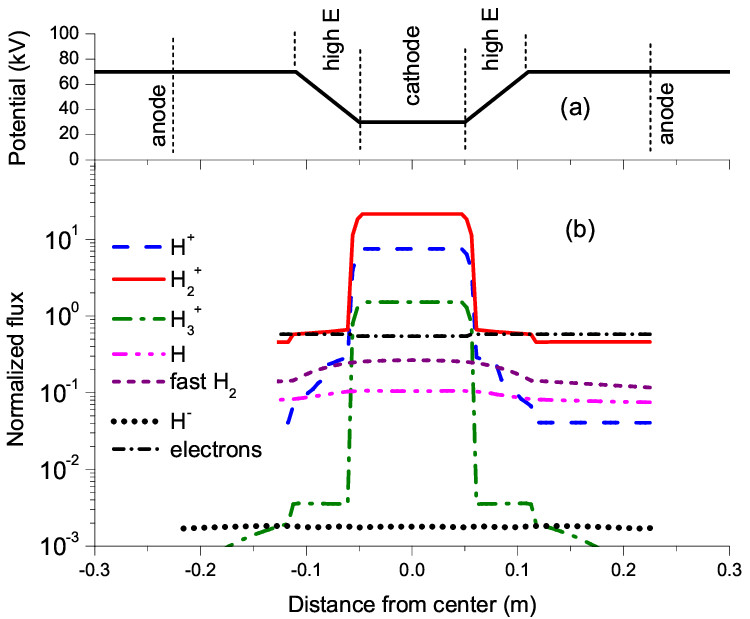}
\end{figure}

\begin{figure}[p]
\includegraphics[width=76mm]{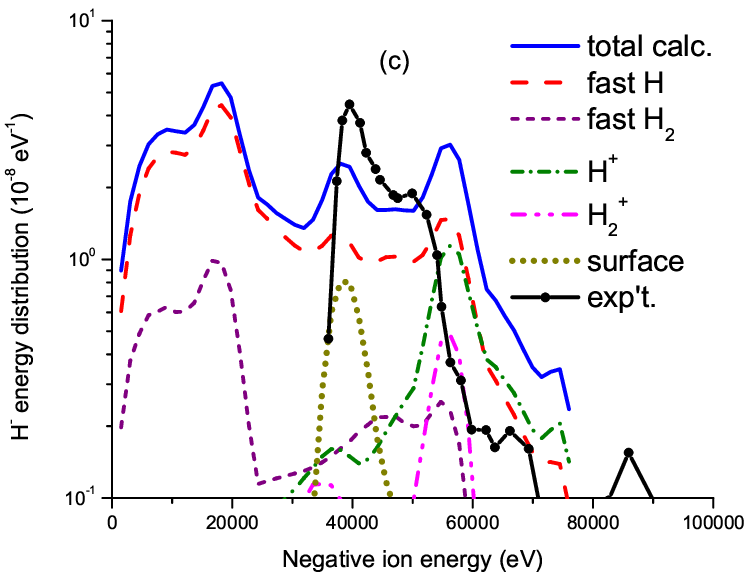}
\end{figure}

\begin{figure}
\caption{\label{fig:boris} (a) Schematic of electric potential as seen by positive ions in the present model of an inertial electrostatic confinement discharge. Vertical dashed lines indicate the positions of the cathode and anode grid wires. (b) Calculated spatial distributions of ions, atoms, and fast molecules for the 70 kV and 2 mTorr experiment of Boris \etal \cite{BOR09}. (c) Comparison of calculated and measured energy distributions for H$^-$ ions. The experimental points are from their figure 4. The smooth curves show the calculated H$^-$ ion energy distribution at the anode grid.}
\end{figure}
\clearpage

%figure 15
\newpage

\begin{figure}[p]
\includegraphics[width=76mm]{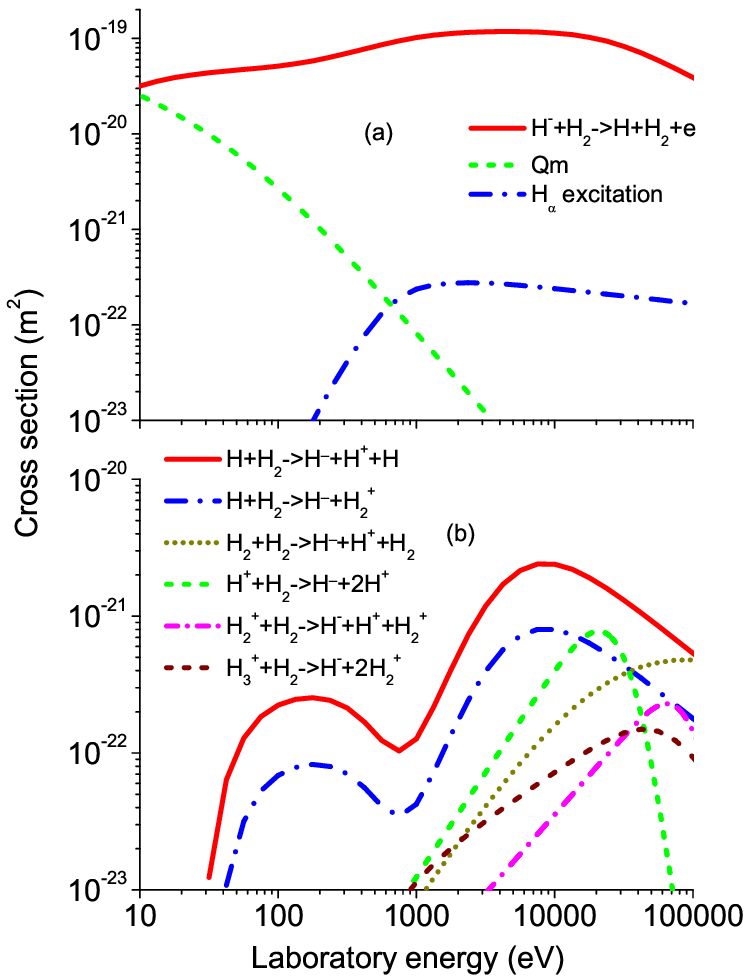}
\end{figure}

\begin{figure}
\caption{\label{fig:negativeionqs} (a) Cross sections for collisions of H$^-$ with H$_2$. (b) Cross sections for H$^-$ production.}
\end{figure}
\clearpage

%figure 16
\newpage
\begin{figure}[p]
\includegraphics[width=76mm]{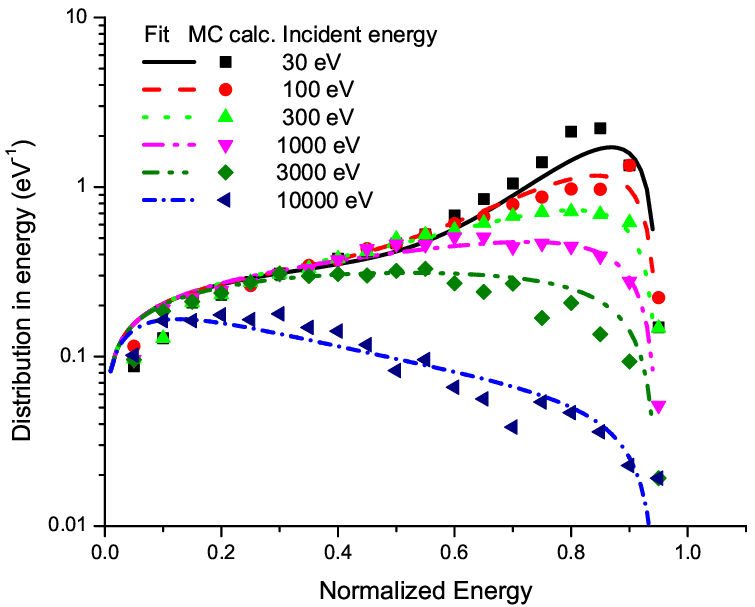}
\end{figure}

\begin{figure}[p]
\includegraphics[width=76mm]{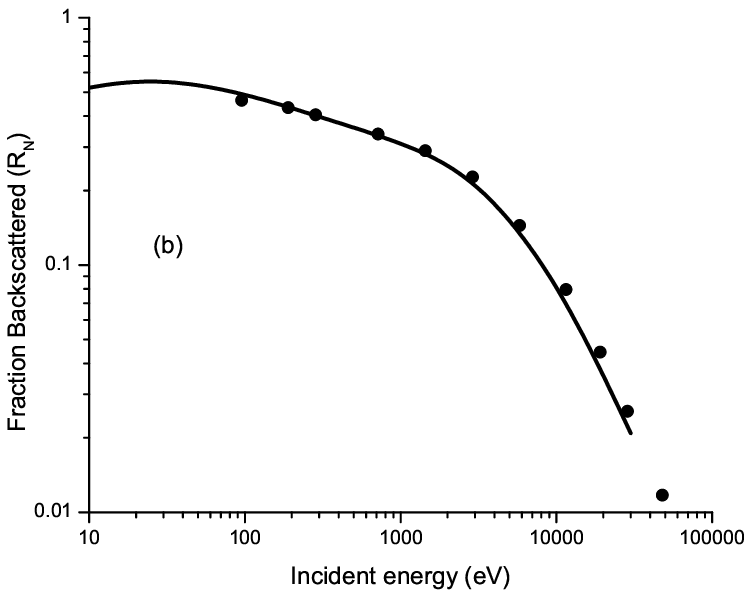}
\end{figure}

\begin{figure}
\caption{\label{fig:backscatter} (a) Energy distribution for backscattered D atoms from nickel bombarded with D$^+$. The points are scaled from the Monte Carlo calculations of Eckstein and Verbeek \cite{ECK84}.  The curves are the present empirical fits to these data. (b) Backscattered fractions obtained from integration of empirical (curve) and Monte Carlo (points) energy distributions for protons on nickel \cite{ECK85}.}
\end{figure}
\clearpage


\begin{thebibliography}{99}
\bibitem{AVP10i} The dispersion width of $w_d = 0.07$ nm (HWHM) corresponds to an exponential decay of an excited state with a lifetime of $\sim 1.6\times 10^{-12}$ s.  This time is much longer than the duration of a typical H + H$_2$ collision or of the resultant H$_2$ dissociation and much shorter than a typical radiative lifetime. Therefore, no particular significance is given to the empirical dispersion profile. The calculated solid curve for $z = 1$ mm shows wings of the dispersion profile at $|\Delta\lambda| > 0.6$ nm that are just below the noise level of the available data.
\bibitem{AVP10j} The model applied to these discharges assumes that the positive ions eventually striking the cylindrical cathode do so at angles and positions such that most of the H atoms they produce are lost to the inner walls of the cathode.  The low-energy electrons emitted from the cathode surface are assumed to follow field lines and enter the external cathode fall. These studies emphasize the need for detailed models of the low pressure hollow cathode, including the trajectories of electrons and positive and negative ions.
\bibitem{AVP10x} Efforts to influence the grid transparency for the lowest energy ions at which the instabily occurs have failed, e.g., changes in the assumed charge transfer cross sections for H$_2^+$ and H$^+$ have no effect. On the other hand, one expects a higher efficiency of positive ion collection by the grid wires for low energy ions.  The enegy bin width for these numerical calculations is typically 1500 eV.
\end{thebibliography}
\end{document}